\documentclass[useAMS,usenatbib]{mn2e}

\def\var{\hbox{V\,393 Sco~}}

\def\msun{\hbox{ $M_{\odot}$~}}

\usepackage{graphicx}
 \usepackage{times}
 \usepackage{lscape}

\title{A study of  the interacting binary V\,393 Scorpii}

\author[ Mennickent et al.]
  {R.E. Mennickent$^{1}$ \thanks{E-mail: rmennick@astro-udec.cl; ESO run 081.D-0457(A)},
  Z. Ko{\l}aczkowski$^{1,2}$, D. Grazcyk$^{1}$ , J. Ojeda$^{1}$   \\
  $^1$Universidad de Concepci\'on, Departamento de F\'{\i}sica,
      Casilla 160-C, Concepci\'on, Chile\\
  $^{2}$  Instytut Astronomiczny Uniwersytetu Wroclawskiego, Kopernika 11, 51-622 Wroclaw, Poland  }
\date{}

\pagerange{\pageref{firstpage}--\pageref{lastpage}} \pubyear{2008}

\def\LaTeX{L\kern-.36em\raise.3ex\hbox{a}\kern-.15em
    T\kern-.1667em\lower.7ex\hbox{E}\kern-.125emX}

\begin{document}

\label{firstpage}

\maketitle

\begin{abstract} 

We present high resolution J-band spectroscopy of V\,393 Sco obtained with the CRIRES at the ESO Paranal Observatory along with a discussion of archival IUE spectra and published broad band magnitudes. The best fit to the spectral energy distribution outside eclipse   gives $T_{1}$= 19000 $\pm$ 500 $K$ for the gainer, $T_{2}$= 7250 $\pm$ 300 $K$ for the donor, $E(B-V)$= 0.13 $\pm$ 0.02 mag. and a distance of $d$= 523 $\pm$ 60 pc, although circumstellar material was not considered in the fit. We argue that V\,393 Sco is not a member of the open cluster M7. The shape of the He\,I 1083 nm line shows orbital  modulations that can be interpreted in terms of an optically thick pseudo-photosphere mimicking a hot B-type star and  relatively large equatorial mass loss  through the Lagrangian L3 point during long cycle minimum. 
IUE spectra show several (usually asymmetric) absorption lines from highly ionized metals and a narrow L$\alpha$ emission core on a broad absorption profile. The overall behavior of these lines suggests the existence of a wind at intermediate latitudes.  
From the analysis of the radial velocities we find $M_{2}/M_{1}$= 0.24 $\pm$ 0.02
and a mass function of $f$= 4.76 $\pm$ 0.24  M$\odot$. 
Our observations favor equatorial mass loss rather than high latitude outflows as the cause for the long variability.
 
\end{abstract}

\begin{keywords}
stars: early-type, stars: evolution, stars: mass-loss, stars: emission-line,
stars: variables-others
\end{keywords}

\section{Introduction}

The star \var (HD 161741; $\alpha_{2000}$ = 17:48:47.603, $\delta_{2000}$= -35:03:25.63, Hipparcos Main Catalogue) was recognized as an Algol-type binary in the Budding (1984) catalogue. An orbital period of 7.71249 days, dominant spectral type of B9 and temperatures of 10.080 $K$ and 6.350 $K$ for the primary and secondary star, are given by Brancewicz  \& Dworak (1980, hereafter BD80). The catalogue of approximate elements of eclipsing binaries by Svechnikov \& Kuznetsova (1990,  hereafter SK90) gives a mass ratio of 0.45 with stellar masses of
3.00 and 1.35 \msun for the primary and secondary  star, respectively. These authors give B9 + [F2] for the stellar components and an inclination of 79.0 degree for the binary. 
 Apart from the above references, the catalog of Budding et al. (2004) quotes $q_{sp}$= 0.12, where $sp$ indicates a mass ratio calculated by using the secondary star radius and the assumption of semi-detached status. These authors also give B3\,III for the primary with a reference to the Hipparcos and Tycho Catalogues. From the above considerations is clear that there are inconsistencies in the basic stellar parameters published for V\,393\,Sco. The star was observed as a  transient radio source by Stewart et al. (1989) with maximum flux density at 8.4 GHz of 9.9 $\pm$ 1.8 mJy. The star is also a 2MASS source (Skrutskie et 
al. 2006) and has been observed by the MSX6C infrared point source catalogue (Egan et al. 2003). \var  is relatively bright, with a reported (variable) $V$ of 
around 7.5 mag. The ephemeris for the main minimum provided by Kreiner (2004) is $T_{o} = 2\,452\,507.7800 + 7.7125772 E$.  V\,393 Sco was briefly discussed by Geraldine Peters in her review on Algol-type binaries (2001).

V\,393 Sco was indicated with additional variability among the eclipsing binaries of the ASAS Catalogue (Pilecki \& Szczygiel 2007). This additional variability is typical of Double Periodic Variables (DPVs, Mennickent et al. 2003, hereafter M03, 2005, 2008, hereafter M08, 2009a, 2009b, 2009c,  hereafter M09), with a period of 253.4 days, about 32.9 times the orbital period. We rapidly recognized \var as a Galactic DPV and determined, from Fourier decomposition of the ASAS $V$ light curve, an ephemeris for the maximum of the long cycle $T_{l} = 2452520.00 + 255.0 * E$.

In our current interpretation of the DPV phenomenon,   the long cycle is interpreted as cycles of mass loss from the binary (M08, Mennickent \& Ko{\l}aczkowski 2009b). This mass loss is revealed, for instance, in the presence of discrete absorption components (DACs)  in Pa$\gamma$ and Pa$\beta$ in the LMC DPV OGLE\,05155332-6925581 (M08).  In this star the radial velocity and strength  of the DACs follow a saw-teeth pattern with the orbital period that M08 interpreted in terms of mass loss from the outer Lagrangian points.  More recently, the long cycles of the Galactic DPV AU\,Mon were interpreted as attenuation due to variable circumbinary material (Desmet et al. 2009).
Our observations of OGLE\,05155332-6925581 motivated us to initiate an observing program looking for these features and similar behavior in the bright Galactic DPV V\,393 Sco,  using the highest spectral resolution possible. At the same time, we also analyzed archival IUE spectra to complement our view of the system.  Brief reports of our investigation have been  published in conference proceedings (M09,   Michalska et al. 2009). In this paper we present results of our spectroscopic investigation of V\,393 Sco at the infrared and ultraviolet spectral ranges, while additional optical spectroscopy  and the study of the light curve will be presented in a separated paper.  This publishing strategy of our rather large observational material has been chosen for the sake of order and paper compactness. Our observations and methodology are summarized in Section 2, our results are given in Section 3, a discussion is presented in Section 4 and we give our conclusions in Section 5.

\section{Observations}

We obtained 27 spectra of \var with the VLT cryogenic high-resolution infrared echelle spectrograph CRIRES  located at the Nasmyth focus A of UT1 (http://www.eso.org/sci/facilities/\\
paranal/instruments/crires/) between august 2008 and may 2009. 22 of these observations were labeled ''completed" and 5 ''executed" by the observing team. These last observations, if not satisfying original demanded atmospheric conditions, 
still are useful for scientific analysis. A summary of our observations is given in Table 1. A slit width of 0.4\arcsec provided a resolving power of  50.000  at the spectral range 1074 --1099 nm. The nearby telluric stars Hip\,087370 (G3V) and Hip\,088154 (G1V) were observed before or after every science exposure, allowing to record the  time dependent atmospheric absorption lines as close as possible (in time and airmass) to our \var spectra. We used the ESO pipeline reduced spectra that are the result of collapsing many  "dithered"  short exposure spectra taken at different slit position in one single spectrum that is finally sky subtracted and wavelength calibrated.
CRIRES has four CCD cameras separated by wavelength gaps, yielding spectral segments CH1 to CH4  corresponding to regions
1074.0--1079.5, 1081.0--1086.5, 1088.0--1093.0, and 1094.5--1099.0 nm, respectively. 

For correction of telluric lines  we used  a variant of the procedure 
described in Maiolino, Rieke \& Rieke (1996).  We built our telluric templates  by dividing the telluric spectra by a synthetic solar-type spectrum interpolated at the same resolution
and wavelength range. We used the NSO/Kitt Peak FTS solar spectrum, produced by NSF/NOAO. 
Then, we used the IRAF\footnote{IRAF is distributed by the National Optical Astronomy Observatories,
 which are operated by the Association of Universities for Research
 in Astronomy, Inc., under cooperative agreement with the National
 Science Foundation.} telluric task to remove telluric absorption lines from the science objects by dividing from every science spectrum a scaled and wavelength centered version of the corresponding telluric spectrum.
This method successfully removed telluric absorption lines from our spectra.  However, we had to deal with an unexpected absorption character present in the telluric and science stars, a relatively wide feature around 1083 nm.
This feature was not present in atlases of telluric absorption lines neither in the ESO simulator of infrared atmospheric transmission. The character seems to be a telluric absorption, since it does not move with the
earth translational motion, even at a target located near the galactic plane as V\,393 Sco (where variability of 40 km/s is expected for the time span of the observations). We discarded a problem with the reduction procedure since the character is present already in the raw data (Fig.\,1).
The complex shape of the feature suggests that it is not a CCD defect. The feature is variable in equivalent width ($EW$), by 30\% per epoch, but deeper in 2009 ($EW$= 0.066 nm) than in 2008 ($EW$= 0.037 nm ). Regarding their nature, interstellar He\,I is expected at this wavelength (Scherb 1968), but with lower intensity and with clear radial velocity variability in the raw data, that is not observed. We speculate that the feature could be He\,I formed in the upper earth atmosphere.

\begin{figure}
\scalebox{1}[1]{\includegraphics[angle=0,width=9cm]{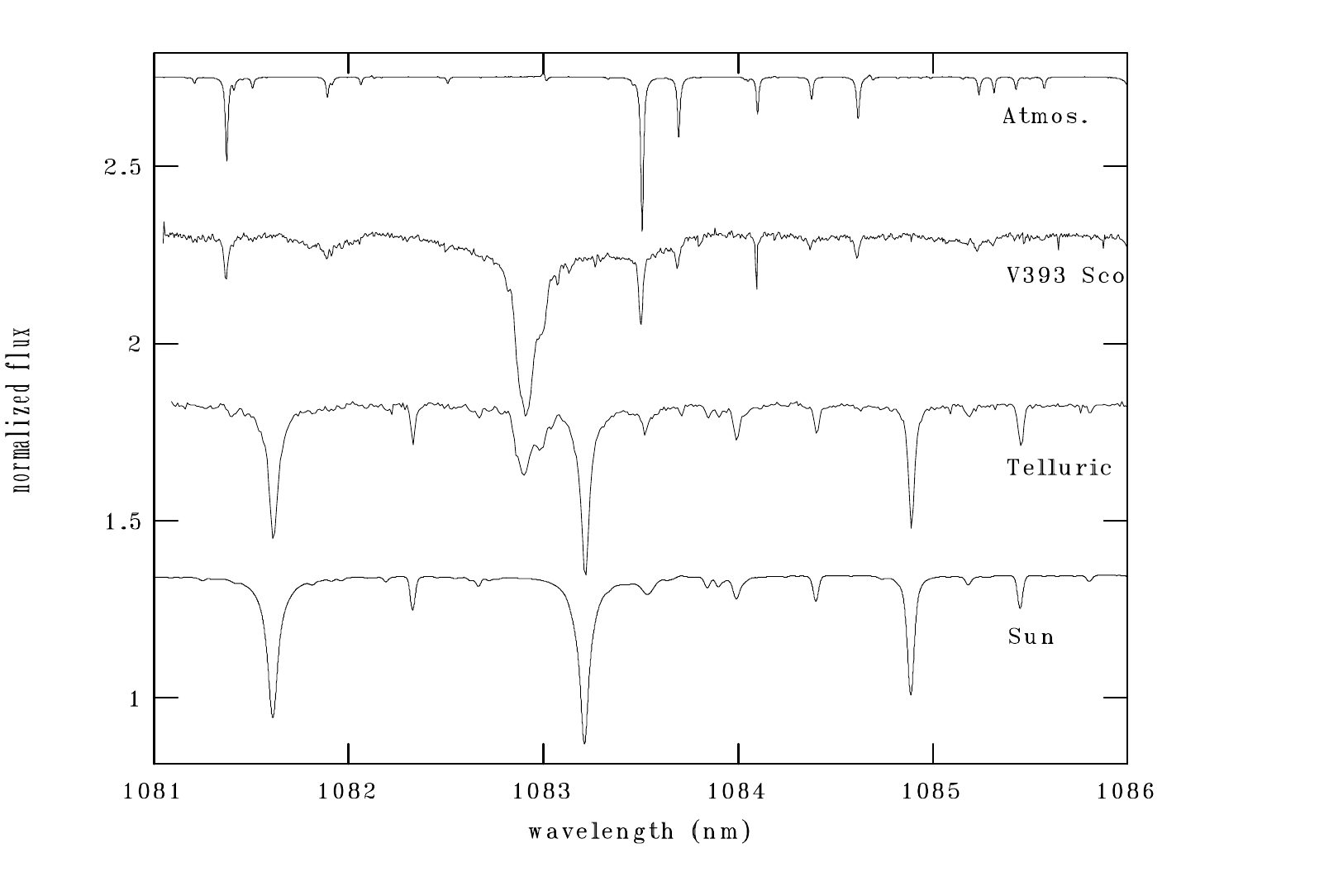}}
\caption{Comparison of raw  telluric and science spectra with the sun spectrum and
atmospheric transmission. The broad feature in the telluric spectrum blueward 1083 nm had to be removed with special algorithms. Its nature is discussed in the text. }
  \label{x}
\end{figure}

Since the conventional method was not optimal for removing completely this feature, 
we developed a special procedure.
We interpolated the continuum above this feature in the telluric spectra (the region is deployed of telluric lines) building a new telluric template, and then corrected only for the classical telluric lines our science objects  in the usual way. This resulted in science objects with removed telluric lines except for the
unknown character. We fit these spectra with high order polynomial excluding the region around the feature and then subtracted the observed spectrum producing the spectrum of the line alone. These line spectra were averaged per epoch producing a model for the line for 2008 and 2009.
We used these models to remove the feature from the science objects using the usual telluric line removal algorithm. The method resulted optimal for line removal. 

For CH3 it was impossible to remove optimally the stellar spectrum from the telluric ones, probably due to small differences in wavelength scales which appear amplified by the high resolving power. For CH3  we cleaned our telluric spectra by removing a fit to the stellar lines made with a deblending algorithm considering a combination of voig profiles of variable depth, width and center. With this procedure we identified and removed 14 stellar lines below 99\% of the continuum intensity level.  Then the remaining "cleaned" telluric spectra were removed  from the science spectra using the usual telluric line removal algorithm. 
Remaining (usually single pixel) outliers were rejected with a rejection algorithm interpolating fluxes  between nearby pixels.

All spectra discussed in this paper are corrected by the earth translational motion. The radial velocities are heliocentric ones.  We use indifferently the words gainer/primary and donor/secondary.

\begin{table*}
\centering
 \caption{Summary of CRIRES infrared spectroscopic observations for V\,393\,Sco. The signal to noise at continuum level is given, along with a quality
 flag (c= observation completed according to requested atmospheric condition, e= observation executed, but not satisfying requested atmospheric conditions). $\phi_{o}$ and $\phi_{l}$ are the orbital and long cycle phases, respectively.
 }
 \begin{tabular}{@{}ccccccc@{}}
 \hline
UT-date (start) &HJD (middle)&$\phi_{o}$& $\phi_{l}$& $S/N_{CH2}$ &$S/N_{CH3}$ &quality\\
\hline
2009-05-14T05:49:05.420 &2454965.75233&     0.6966 &0.5912&65&53&c\\
2009-05-14T05:01:05.661 &2454965.71900&     0.6923 &0.5911&200&140&e\\ 
2009-04-25T09:54:11.477 &2454946.92128&     0.2550 &0.5173&120&100&c\\
2009-04-24T08:23:37.979 &2454945.85831&     0.1172 &0.5132&124&114&c\\
2009-04-23T08:01:08.944 &2454944.84262&     0.9855 &0.5092&80&70&c\\
2009-04-22T08:21:52.964 &2454943.85694&     0.8577 &0.5053&130&101&c\\
2009-04-21T08:12:30.627 &2454942.85035&     0.7272 &0.5014&150 &111&c\\
2009-04-20T07:05:52.433 &2454941.80399&     0.5915 &0.4973&100&78&c\\
2009-04-19T07:13:06.427 &2454940.80893&     0.4625 &0.4934&80&67&c\\
2008-09-30T23:41:10.294 &2454740.49107& 0.4906& 0.7078 &115    &98&c\\
2008-09-30T00:06:44.212 &2454739.50892& 0.3633& 0.7040 &87    &92&c\\
2008-09-29T23:56:54.473 &2454739.50210& 0.3624& 0.7039 &41    &40&e\\
2008-09-29T00:28:13.877 &2454738.52395& 0.2356& 0.7001 &52    &58&e\\
2008-09-25T00:24:54.042 &2454734.52202& 0.7167& 0.6844 &116    &82&c\\
2008-09-24T00:17:12.919 &2454733.51678& 0.5864& 0.6805 &175    &132&c\\
2008-09-22T23:55:20.715 &2454732.50168& 0.4548& 0.6765 &120    &150&c\\
2008-08-28T23:03:49.776 &2454707.46829& 0.2090& 0.5783 &130   &128&c\\
2008-08-25T23:12:18.433 &2454704.47444& 0.8208& 0.5666 &130   &107&c\\
2008-08-24T23:45:02.189 &2454703.49725& 0.6941& 0.5627 &160   &82&c\\
2008-08-24T01:16:15.657 &2454702.56068& 0.5726& 0.5591 &90   &121&c\\
2008-08-22T02:46:27.071 &2454700.62348& 0.3215& 0.5514 &150   &118&c\\
2008-08-21T02:26:25.243 &2454699.60967& 0.1900& 0.5475 &80   &107&c\\
2008-08-20T03:45:45.163 &2454698.66483& 0.0675& 0.5438 &79  &67&c\\

2008-08-19T03:52:02.058 &2454697.66927& 0.9384& 0.5399 &35  &36&c\\
2008-08-19T03:57:04.829 &2454697.67277& 0.9389& 0.5399 &160  &129&c\\
2008-08-17T04:54:27.220 &2454695.71278& 0.6847& 0.5322 &70   &62 &e\\
2008-08-16T02:40:47.850 &2454694.62005& 0.5431& 0.5279 &131  &121 &c\\
\hline
\end{tabular}
\end{table*}

\section{Results}

\subsection{Modeling the spectral energy distribution: stellar parameters, distance and reddening}

\begin{figure}
\scalebox{1}[1]{\includegraphics[angle=0,width=8cm]{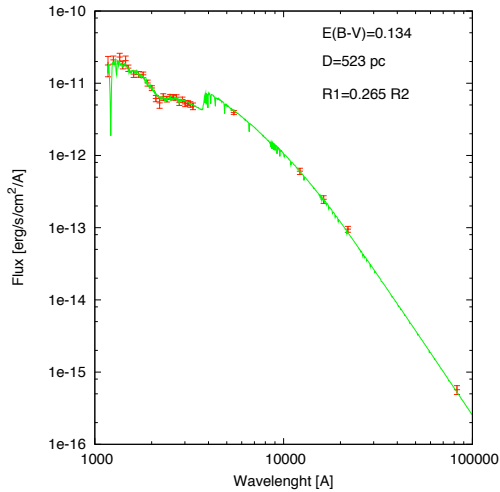}}
\caption{The SED with the fluxes derived from published broad-band photometry and IUE flux calibrated spectra along with the best fit. Some derived parameters are shown.}
  \label{x}
\end{figure}

We compiled  fluxes at different wavelengths  from several sources to build the spectral energy distribution (SED) of V\,393 Sco. Magnitudes $m_{\lambda}$ were transformed to fluxes using standard zero magnitude fluxes.  The result of our compilation is shown in Table 2. To these fluxes we added UV fluxes derived from the spectrophotometry described in Section 3.2, averaging fluxes outside eclipse at several small wavelength ranges.  The sample of all these fluxes is analyzed in this section, with the condition  that the observations correspond to phases outside eclipse,  except the point at $\lambda$ 8280 nm that was included without knowing if this condition was met. The last flux in Table 2 corresponds to the mean of several 8.4 GHz flares reported by Stewart et al. (1989), who attributed this radio emission to gyrosynchrotron radiation from  mildly relativistic electrons trapped in magnetic fields around one of the  stars. Due to the transient nature of this detection, this datapoint will not be considered in our analysis. In addition, we rejected the outlier at 4290 nm for not following the rather clear general tendency. The remaining fluxes are shown in Fig.\,2.

We performed a fit to the SED by means of the Marquant-Levenberg non-linear least square algorithm by  minimization of $\chi^{2}$ of the function:\\

\begin{small}
$f_{\lambda}=  f_{\lambda,0} 10^{-0.4E(B-V)[k(\lambda -V)+R(V)]} $\hfill(1)\\
\end{small}

\noindent
where:\\

\begin{small}
$ f_{\lambda,0}= (R_{2}/d)^2 [(R_{1}/R_{2})^{2} f_{1, \lambda} +  f_{2, \lambda}] $\hfill(2)\\
\end{small}

\noindent 
and  $f_{1}$ and $f_{2}$ are the fluxes of the primary and secondary star,  $k(\lambda-V) \equiv E(\lambda-V)/E(B-V)$ is the normalized extinction
curve, $R(V) \equiv A(\lambda)/E(B-V)$ is the ratio of reddening to extinction at $V$, $d$ is the distance to the binary, $R_{2}$ is secondary physical radius and $R_{1}/R_{2}$ is the ratio of the primary radius to the secondary radius. 
In absence of any other better approach, we used the average Galactic Extinction Curve parametrized by Fitzpatrick \& Massa (2007) to calculate reddened fluxes. The stellar fluxes were taken from the grid of models given by Castelli \& Kurucz (http://wwwuser.oat.ts.astro.it/castelli/). We used models with solar chemical abundance. The scale of the system was set according to the orbital solution with mass ratio q= 0.24 (see section 4.1). Assuming that secondary fills its Roche-lobe we set secondary radius to $R_{2}$= 9.2 $R_{\odot}$. The free parameters of the fitting were $d$, $E(B-V)$ and $R_{1}/R_{2}$ for each chosen pair of synthetic stellar fluxes. 


The best fit, minimizing $\chi^{2}$,  gave $T_{1}$= 19000 $\pm$ 500 K, $T_{2}$= 7250 $\pm$ 300 K, $\log g_{1}$= 4.5 $\pm$ 0.3, $\log g_{2}$= 3.0 $\pm$ 0.3, $d$= 520 $\pm$ 60 pc, $E(B-V)$= 0.13 $\pm$ 0.02 and $R_{1}/R_{2}$= 0.27 $\pm$ 0.03 and it is also shown in Fig.\,2.
The formal errors were taken from an inspection of the parameters around the $\chi^{2}$ minimum. 
Although the fit appearance looks fine, we must to remember that V\,393 Sco is not a simple binary, but show emission lines indicating  the presence of circumstellar matter, that was not included in the fit function. This fact, along with the possible perturbation introduced by the system intrinsic variability, long and short photometric cycles, force us to be careful about the significance of the above solution.
 In particular, the above arguments should help to explain the difference observed with the donor temperature of 7900 $\pm$ 100 $K$ derived from the fit to selected optical spectral lines (M09).

V\,393 Sco is projected against the open galactic cluster M7 (log T = 8.475,  d= 301 pc, E(B-V)= 0.103 mag; http://www.univie.ac.at/webda/).  However, \var 
shows  different proper motion, color and parallax when compared with those of well established NGC\,6475  stars,  to be considered a probable member (Fig.\,3).  Our distance $d$= 520 pc and color excess $E(B-V)$= 0.13 confirm this conclusion.

\begin{table}
\centering
 \caption{A compilation of fluxes derived from magnitudes reported in the literature. Uncertainties are quoted when available. Fluxes are given in erg cm$^{-2}$s$^{-1}$A$^{-1}$ except when indicated. Only fluxes outside eclipse and that for 8280 nm were considered for the SED fit.}
 \begin{tabular}{@{}ccccc@{}}
 \hline   
 $\lambda$ (nm) & $f_{\lambda}$ & $\Phi_{o},\Phi_{l}$ & Reference\\
\hline
156.5 &1.58(05)E-11    &NA           &Thompson et al. 1978\\
196.5  &1.00(08)E-11 &NA &Thompson et al. 1978\\
236.5  &7.50(54)E-12 &NA&Thompson et al. 1978\\
274      &5.48(15)E-12  &NA&Thompson et al. 1978\\
545     & 4.51E-12     &NA&Thompson 1978\\
545	   &     3.57E-12  &0.35,NA	        &M07\\
545     & 3.18E-12                &0.44,0.39    &Hauck et al. 1997 \\   
1220&	6.05(11)E-13&0.35,0.19	&Skrutskie et 
al. 2006\\
1630&	2.47(09)E-13&0.35,0.19	&Skrutskie et 
al. 2006\\
2190&	9.53(16)E-14&0.35,0.19	&Skrutskie et 
al. 2006\\
4290 &	3.29(1.34)E-13&NA	&Egan et al. 2003\\
8280 &	5.70(43)E-16&NA	&Egan et al. 2003\\
35.7E6 &5.9mJy &NA &Steward et al.  1989\\
\hline
\end{tabular}
\end{table}

\subsection{The UV spectral energy distribution}

We found 10 flux calibrated spectra of V\,393 Sco in the IUE archive (http://archive.stsci.edu/iue). Three of them are low resolution spectra ($R$= 6 \AA) and 7 high resolution spectra ($R$= 0.2 \AA),
2 are in the region 184.5-310.5 nm and 8 in the region 115.0-193.0 nm , all of them were taken with the large (10" $\times$ 20" ) aperture (Table 3). 

We identified
several stellar absorption lines of Si\,II-III-IV, Al\,II-III, C\,II-III-IV, N\,V and  Mg\,II. Some of these lines show narrow interstellar cores, especially some Si\,II lines.  These interstellar lines are easily identified when compared with the broader absorption lines formed in the binary and provides a measure of the stability and accuracy of the wavelength calibration ($\pm$ 7 km/s).  We compared our normalized spectra with the grid of reference IUE spectra taken at the same resolution by  Rountree \& Sonneborn (1991), deriving a spectral type  for the primary, based on the appearance of specific lines, of B2-B3, in agreement with the parameters determined in Section 3.1  and also with one of the spectral types for the primary quoted by Budding et al. (2004). The strength of Si\,IV 140, Al\,III 185 and the weakness of C\,IV 155 and almost absence of He\,II lines are the dominant indicators for establishing the spectral type.

\subsection{The UV spectroscopic variability: evidence for outflows}

We measured fluxes in the continuum using three 20-nm width ''squared" bandpasses centered at 140, 160 and 180 nm  (Table 3). The orbital behavior suggests small fluxes during eclipses and large near quadratures, with small color changes (Fig.\,4). The amplitude of variability amounts to 0.18 magnitudes.  

The spectra show strong and broad L$\alpha$ absorption with a small and narrow  emission core (FWHM $\sim$ 100 km/s). The L$\alpha$ emission strength is higher at phase 0.55 and subsequently seems to recover the ''plateau" value at phase 0.83 (Fig.\,4). This variability cannot be caused by
changes in the continuum level, and probably reveal changes in the visibility of the forming region through the orbital cycle.









\begin{figure}
\scalebox{1}[1]{\includegraphics[angle=0,width=8.5cm]{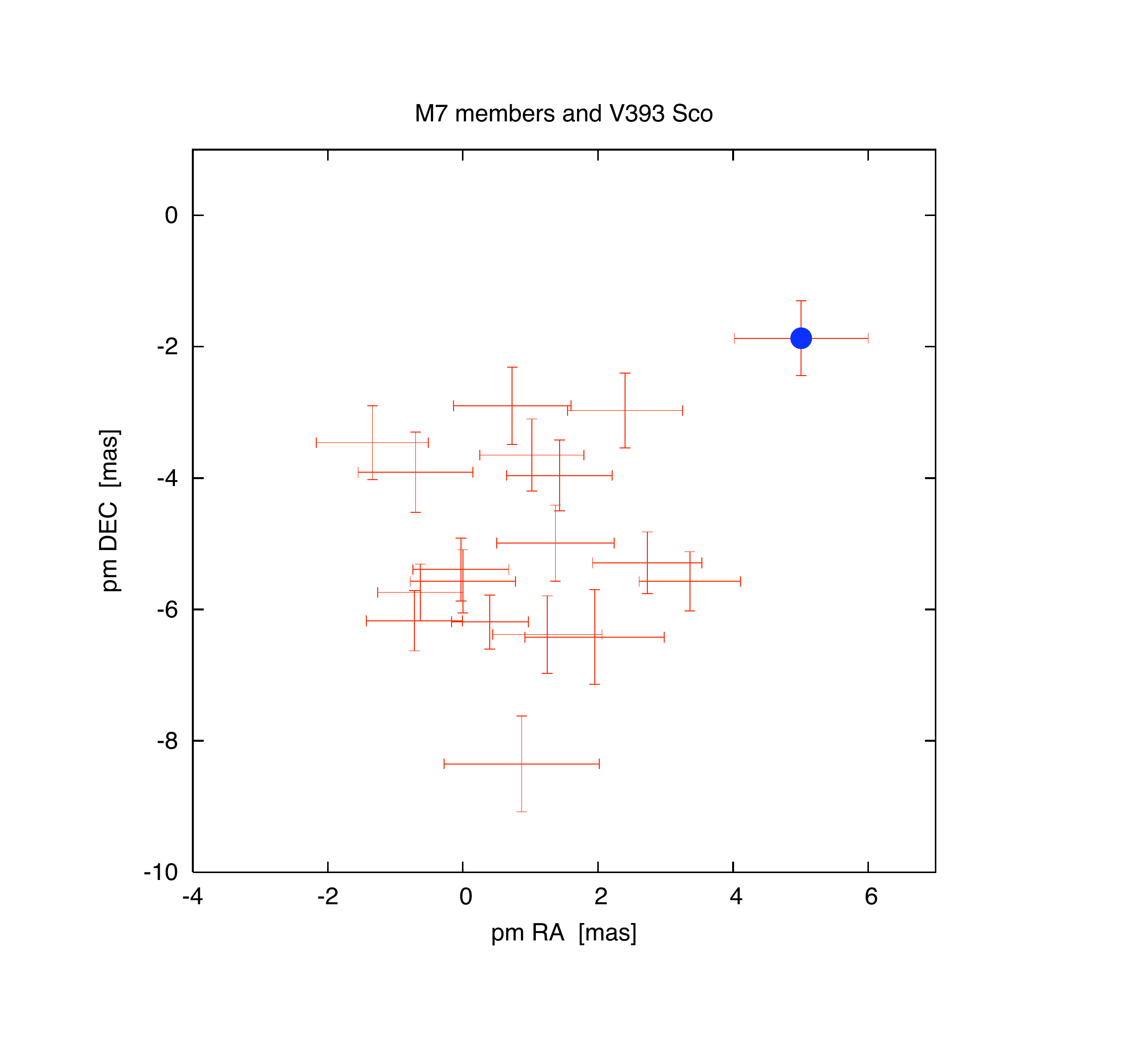}}\\
\scalebox{1}[1]{\includegraphics[angle=0,width=8.5cm]{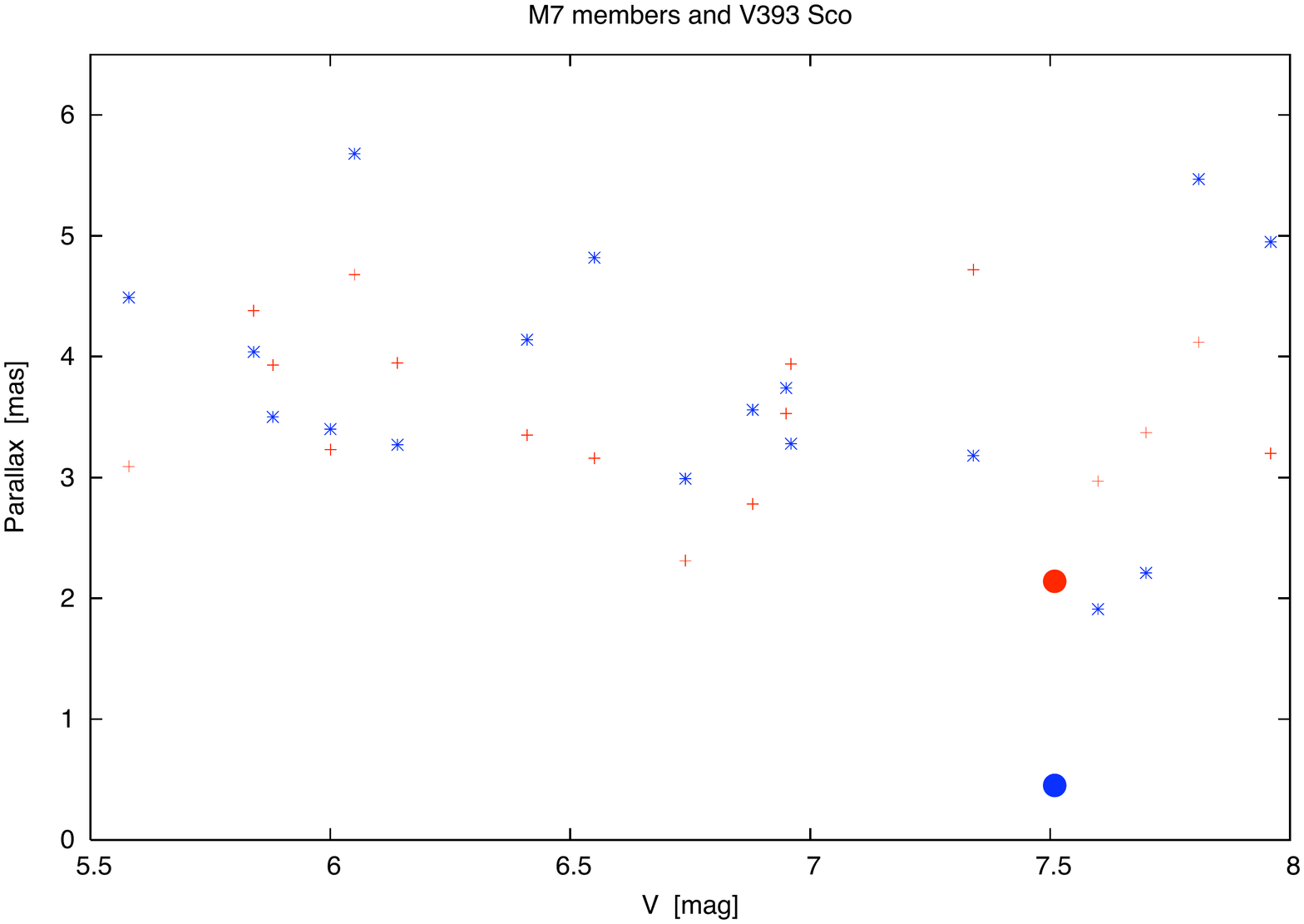}}\\
\scalebox{1}[1]{\includegraphics[angle=0,width=8.5cm]{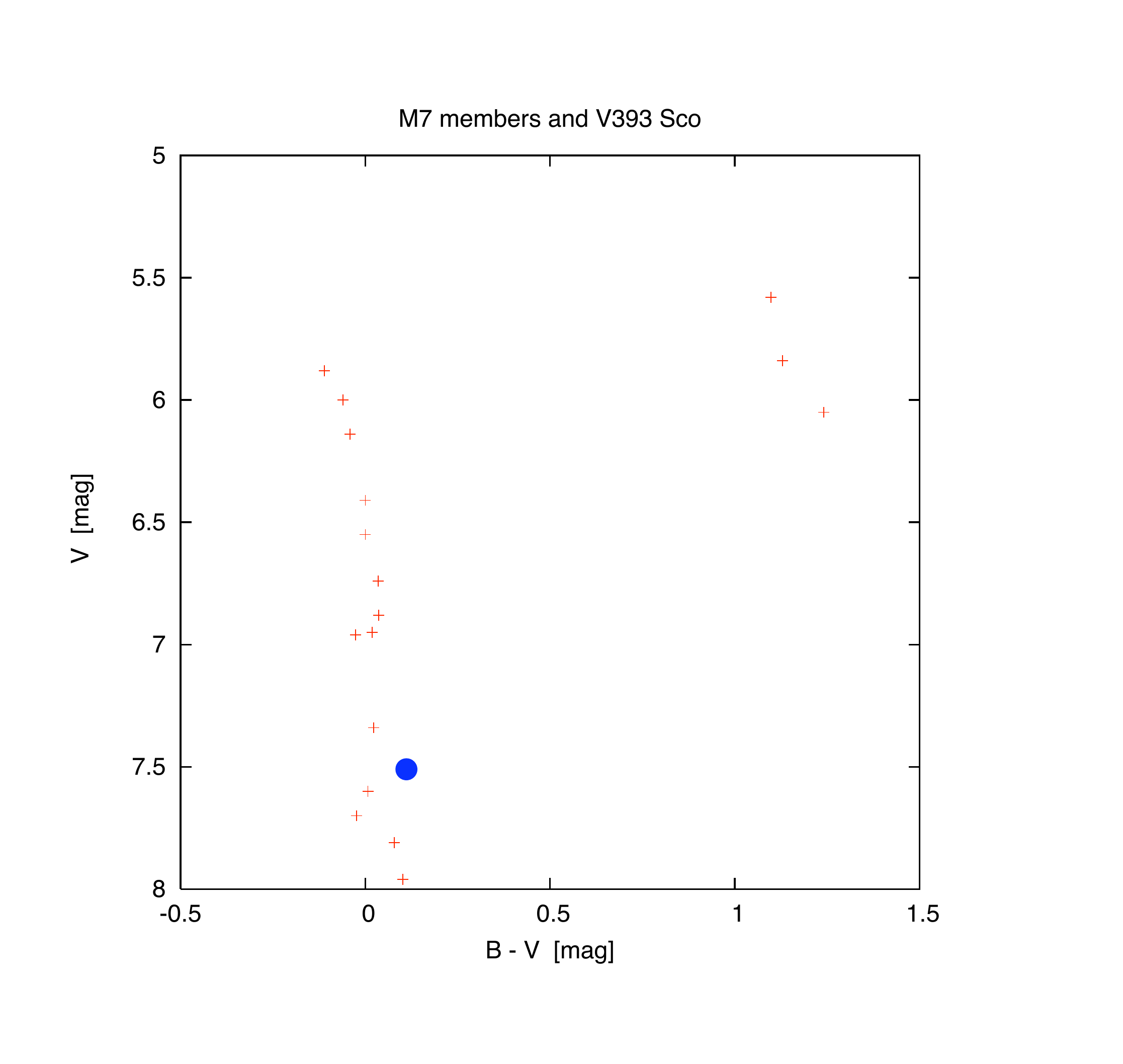}}
\caption{\var (big dot) compared with members of the cluster M7. Parallaxes and proper motions are from the Hipparcos Catalogue, second edition. In the middle frame we also include first edition data (pluses and the upper big dot). }
  \label{x}
\end{figure}


There are clear indications of mass loss, revealed in the presence of
depressed blue wings in  many lines, especially Si\,IV lines.
Si\,IV lines are strongly asymmetric, except at orbital 
phase 0.55, when become of triangular shape. 
N\,V 123.88 is notable in showing a strong modulation of $FWHM$ with orbital phase, being larger
at phase 0.71 and weaker and narrower at phase 0.21 (Fig.\,5).
Lines of lower ionization ions like C\,II and Al\,III usually show weaker asymmetry than Si\,IV lines.

\begin{figure}
\scalebox{1}[1]{\includegraphics[angle=0,width=8cm]{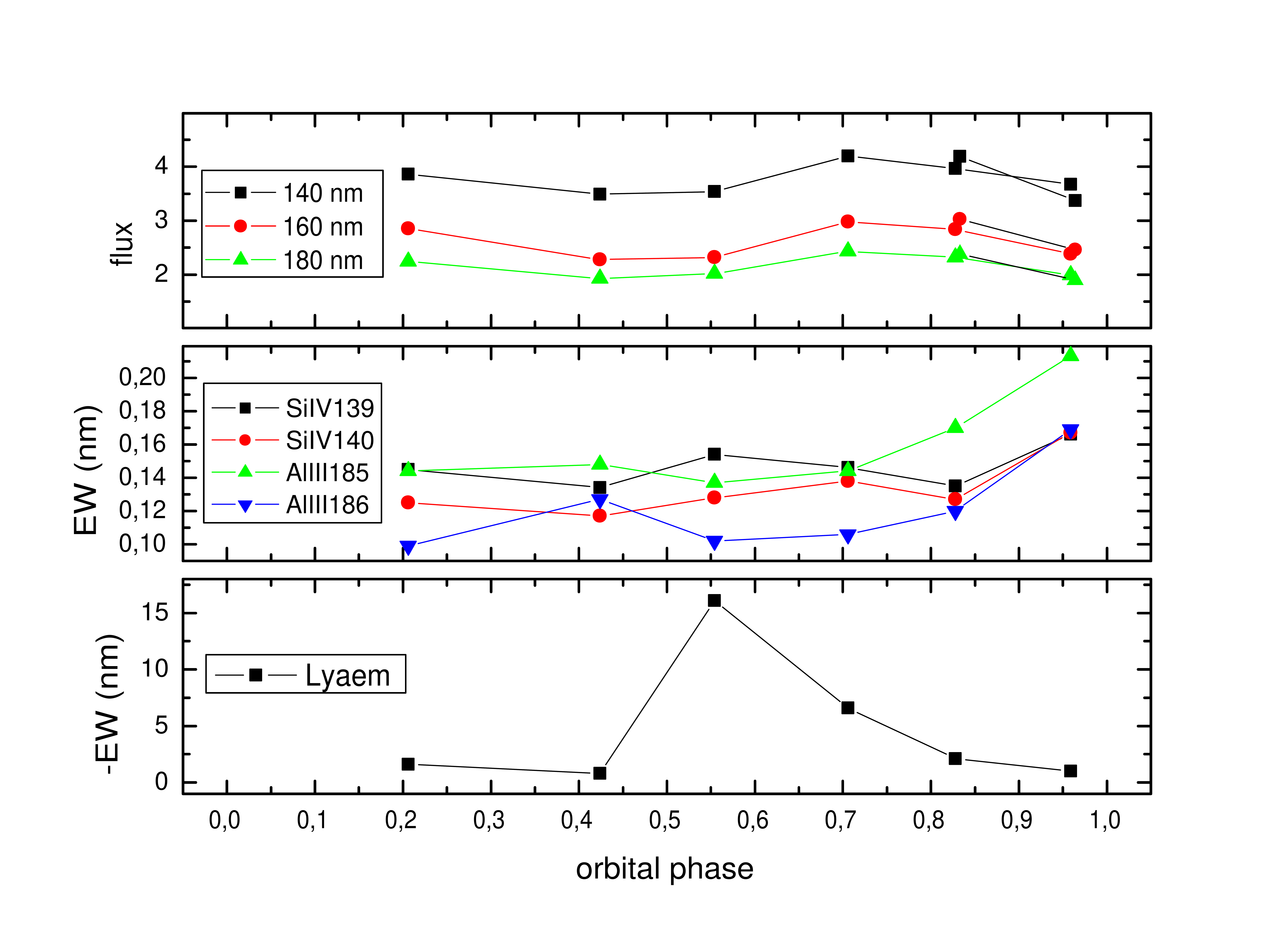}}
\caption{Full equivalent widths and fluxes versus $\Phi_{o}$. The 2-point disconnected lines
in the upper graph correspond to low resolution spectrum data.}
  \label{x}
\end{figure}

\begin{figure*}
\scalebox{1}[1]{\includegraphics[angle=0,width=8cm]{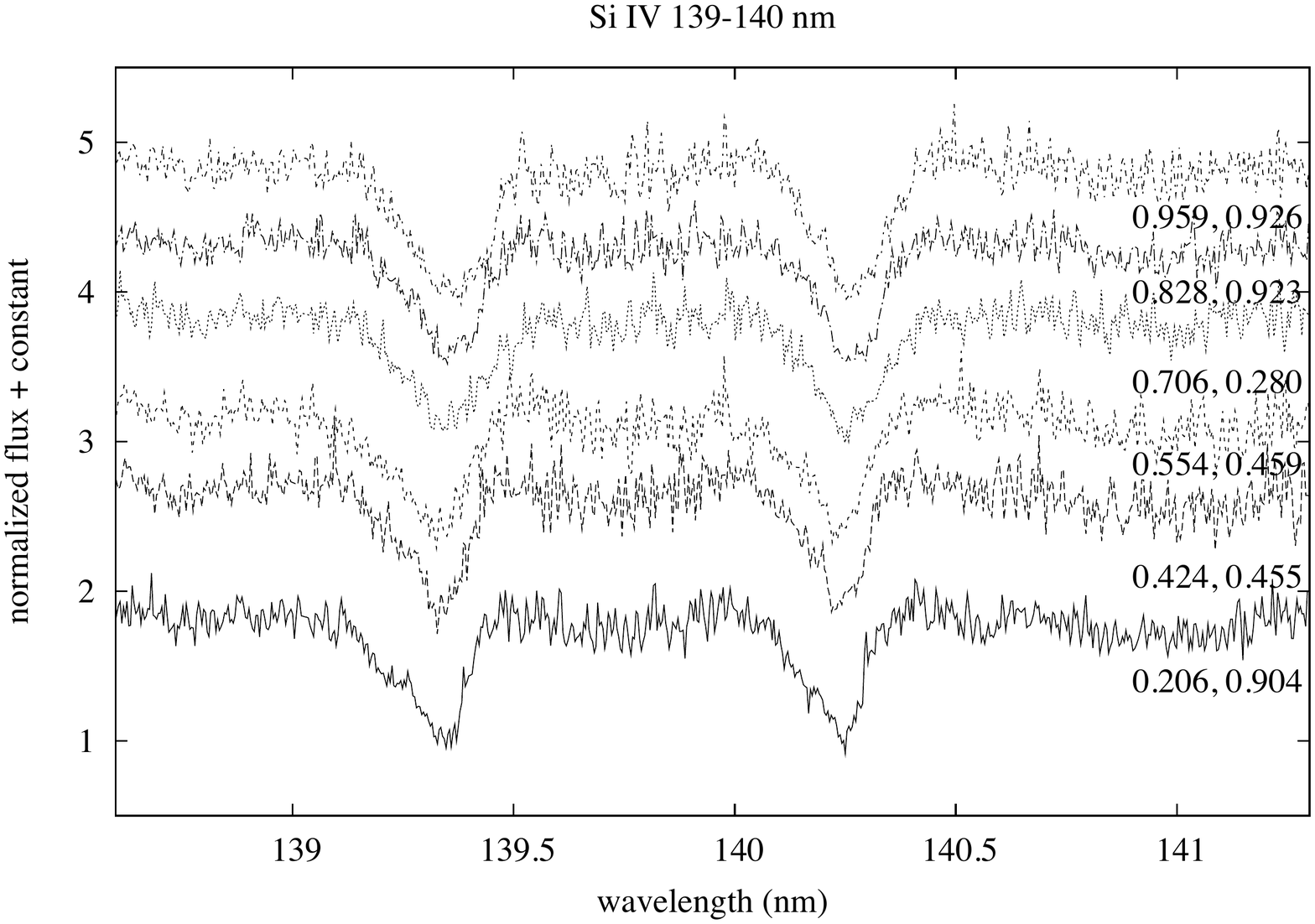}}\\
\scalebox{1}[1]{\includegraphics[angle=0,width=8cm]{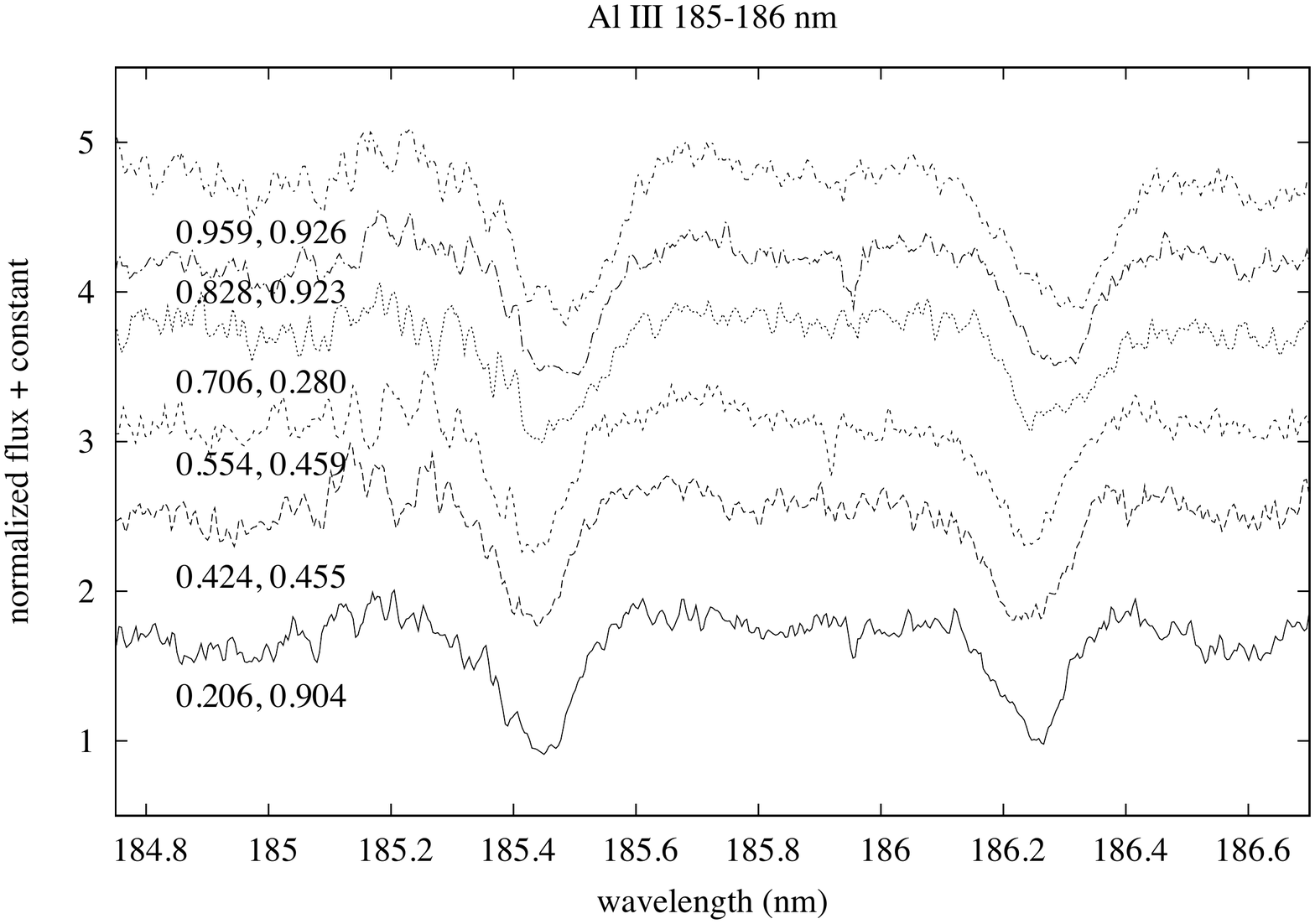}}\\
\scalebox{1}[1]{\includegraphics[angle=0,width=8cm]{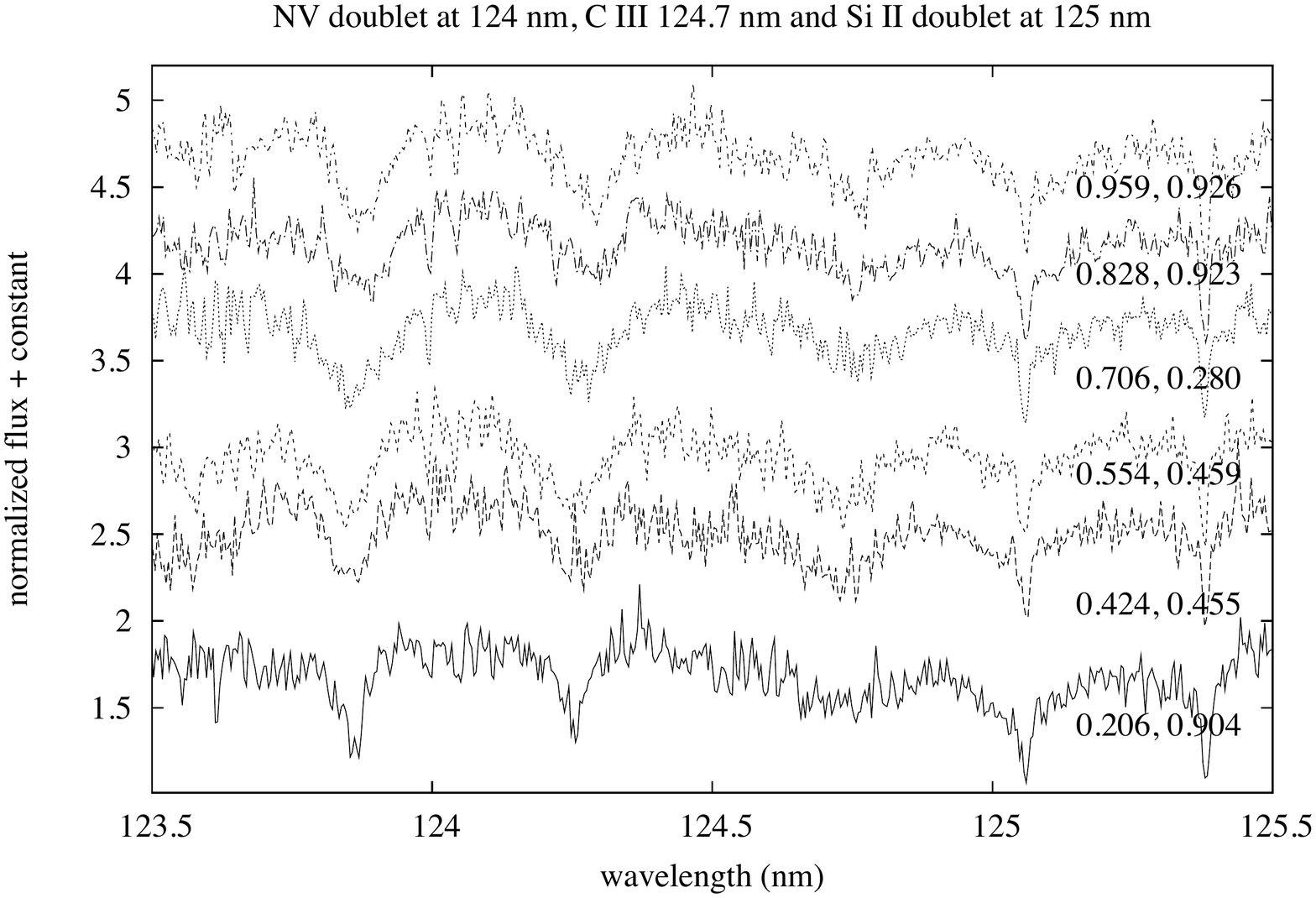}}
\caption{ The behavior of Si\,IV, Al\,III and N\,V lines. Orbital and long cycle phases are labeled for each spectrum.}
  \label{x}
\end{figure*}

We measured radial velocities, widths and equivalent widths for some prominent lines. For some asymmetrical lines we used deblending routines to separate the line in a blue (C1) and red (C2) gaussian component.  Measurements are given in Table 4. Equivalent widths of these lines increase near main eclipse, which could indicate a high latitude origin  (Fig.\,4). 
The blue component disappears at phase 0.7, but the fact that the total $EW$ remains constant at this phase suggests that the blue component is inside the red component profile, being difficult to measure with our deblending routine. The 
velocities of the line main components describe low amplitude modulations roughly in antiphase with the velocities of the donor, but with a phase shift;  they crosses the mean value approximately at phase 0.65 (Fig.\,6). The same behavior is observed for the L$\alpha$ emission and  blue components, but these later show higher amplitude variability.  All lines show  blue shifted $\gamma$ velocities, regarding the systemic velocity inferred from infrared lines, viz\, -1.7 $\pm$ 2 km/s (see next section), again consistent with global mass outflows. Results for sinus fits to the RVs are given in Table 5. The fit for Si\,IV is degraded regarding the fits to other silicon and aluminium lines by the outlier at phase 0.96. 

The phasing and small amplitude  of the RVs of the high ionization absorption lines and L$\alpha$ emission are consistent  with an origin in a region around the more massive star. Another interesting feature is the large $\gamma$ shift  observed in superionized UV lines (especially the blue components). This is interpreted as evidence of line absorption in an expanding media. On the contrary,  the $L\alpha$ emission $\gamma$ shift is so small, and the line so narrow, that is probably formed in a very localized region near the primary, some kind of interacting region between the gas stream and the circumprimary envelope. 

We conclude that very likely the observed IUE absorption lines are formed in a complex hot region around the primary.  This region is characterized by high latitude outflows and could form a kind of expanding pseudo-photosphere for the primary star.

\begin{table*}
\centering
 \caption{Summary of archival  IUE spectra analyzed in this paper. We give the HJD (middle, -244\,0000) and  
the average dereddened flux (in 10$^{-11}$ ergs s$^{-1}$ cm$^{-2}$ \AA$^{-1}$) in 20-nm width "squared" bandpasses  centered at 140, 160 and 180 nm.}
 \begin{tabular}{@{}cccccccccc@{}}
 \hline
Spectrum   &$\Delta \lambda$ (nm) &resolution &HJD &UT-Date & exptime (s) &$\phi_{o}$ &$\phi_{l}$  & S/N &flux  \\  
\hline   
SWP36222&115.0-193.0&low & 7655.28398 & 08/05/89&  38    &0.833   &0.923  &13    &4.19/3.03/2.38   \\
LWP15481&184.5-310.5&low & 7655.28768 & 08/05/89&  24    &0.833   &0.923  &9       & - \\
SWP36232&115.0-193.0&low& 7656.29371 & 09/05/89 & 46     &0.964   &0.927  &13    &3.37/2.46//1.90  \\
LWP15447&184.5-310.5&high& 7650.41954 & 03/05/89 & 900 &0.203   &0.904  &10     &- \\
SWP36189&115.0-193.0&high& 7650.44167& 03/05/89& 2340&0.206   &0.904  &8      &3.86/2.85/2.24\\
SWP36221&115.0-193.0&high& 7655.24279& 08/05/89& 3300&0.828   &0.923  &9      &3.97/2.84/2.32\\
SWP36231&115.0-193.0&high& 7656.25318& 09/05/89& 3600&0.959   &0.926  &8      &3.67/2.38/1.99\\
SWP37117&115.0-193.0&high& 7790.95742& 21/09/89& 3300&0.424   &0.455  &7      &3.49/2.28/1.93\\
SWP37145&115.0-193.0&high& 7791.95566& 22/09/89& 3300&0.554   &0.459  &6      &3.54/2.32/2.02\\
SWP38632&115.0-193.0&high& 8001.36726& 19/04/90& 2700&0.706   &0.280  &9      &4.20/2.98/2.43\\
\hline
\end{tabular}
\end{table*}

\begin{table*}
\centering
 \caption{Heliocentric radial velocities (km/s), $FWHM$ (km/s, in parenthesis) and equivalent widths (nm, after the semicolon) for selected lines. C1 and C2 refer to 
 components of double gaussian fits.}
 \begin{tabular}{@{}cccccccc@{}}
 \hline   
Spectrum     &L$\alpha$-em & Si\,IV 139.37 C1 & Si\,IV 139.37 C2 &Si\,IV 140.28 C1 &Si\,IV 140.28  C2 &Al\,III 185.47 & Al\,III 186.28 \\
\hline
SWP36189 & 29(100);-1.6  &-318(281);0.058    &-69(220);0.087  &-296(233);0.044&-83(210);0.081 &-80(247);0.144&-72(190);0.099 \\
SWP36221 & 24(94);-2.1    &-323(178);0.017    &-44(312);0.118  &-294(267);0.016 &-42(281);0.111 &-23(297);0.170 &-10(222);0.120\\
SWP36231 & 28(93);-1.0    &-125(343);0.120    &72(205);0.046   &-238(280);0.037 &-22(329);0.130 &-23(333);0.213 &0(275);0.169\\
SWP37117 & -8(90);-0.8     &-337(227);0.034    &-80(250);0.100  &-294(246);0.034 &-85(221);0.083 &-99(249);0.148 &-91(217);0.127\\
SWP37145 & -10(92);-16.1&-404(331);0.045    &-94(328);0.109  &-355(170);0.023 &-101(267);0.105 &-96(225);0.137 &-85(185);0.102\\
SWP38632 & 38(98);-6.6    &                                 &  -70(432);0.146                            &    &-65(379);0.138  &-43(276);0.144  &-19(238);0.106\\
\hline
\end{tabular}
\end{table*}


\subsection{Infrared spectroscopy: signatures of mass transfer and mass loss}

The segment CH1 shows a noisy spectrum  without spectral features while CH4 shows the red wing of Pa$\gamma$ without any other major spectral feature. The center of Pa$\gamma$  drops just in the gap between CH3 and CH4;  no available  grating combination allows recording the whole Pa$\gamma$ line. Here we report  our observations for the spectral segments CH2 and CH3.

CH2 displays the variable He\,I 1083.306 nm line along with weak Mg\,I 1081.408 nm
and Si\,I 1084.681 nm. For line identification we consulted Wallace et al. (2000).  CH3 displays He\,I 1091.595 nm, Mg\,II 1091.727 nm and possibly Si\,I 1088.826 nm. 
We did not observe discrete absorption components blueward $250$ km/s of the  Pa$\gamma$ rest wavelength,  contrary to the DPV OGLE\,05144332-6925581 (M08). All spectra were normalized to the continuum before performing our RV measurements and analysis, maintaining the depression of the Pa$\gamma$ blue absorption wing 
in the CH3 spectra. The variability He\,I 1083 and Mg\,II 1092 is illustrated in Fig.\,7 and analyzed in the forthcoming paragraphs.

The line Mg\,II 1091.7 is sometimes blended with the Pa$\gamma$ blue wing and He\,I 1091.595,  so a deblending method was needed to isolate it for analysis. We performed this task with the IRAF deblending "d" routine available in the "splot" package. This resulted in a series of positions for the line center that were used to calculate the radial velocities. 
These RVs indicate a circular orbit (Fig.\,8), so a sinus function was selected to fit them,  
obtaining $K_{donor}$= 181.1 $\pm$ 2.8 km/s and $\gamma_{donor}$= -1.7 $\pm$ 2.1 km/s with $rms$ 10 km/s (Fig.\,9). The rather large $rms$ possibly reflects the uncertainties in working with a heavily blended line.   A sinus fit to the few RVs possible to measure for the
He\,I 1091.595 line gives $K_{1}$= 42  $\pm$ 7 km/s (similar to the value found for the superionized lines) and $\gamma$ = 16 $\pm$ 6 km/s (much larger). The  weak emission peaks surrounding the absorption line Mg\,II 1091.7 around $\Phi_{o}$= 0.20-0.30 are likely the effect of blending with He\,I 1091.6 and Pa$\gamma$. 

We measured the equivalent width of Mg\,II 1091.6 finding an orbital modulation that can be represented by:\\

\begin{small}
$EW_{1091}= 0.046(2) + 0.019(3) \times \sin[2\pi(\Phi_{o}+0.23(2))]$\hfill(3)\\
\end{small}

\noindent with $rms$ 0.009 nm. The maximum strength during main eclipse and minimum strength during secondary eclipse is just the expected behavior  for the line formed in the donor.



He\,I 1083.306  shows a complex pattern of variability during the orbital cycle: it is narrower near main eclipse,  wider at $\Phi_{o}$=  0.5 and show a moving  feature mostly visible at quadratures, especially around phase 0.75. This feature consists on a discrete emission component and a discrete absorption component  (Fig.\,7). At first glance is not clear if the emission produces the absorption or viceversa. The emission feature and the associated absorption appear at negative and positive velocities at each half of the orbital cycle, not like DACs observed in the Paschen lines of OGLE\,05144332-692558 that are limited to the blue wing profile only (M08). Their RVs closely follow the RV of the donor (Fig.\,8 and Table 5). We give two arguments favoring the reality of the absorption feature (i.e. the emission is the artifact produced by the absorption): (1) the "decoupling" of RVs around phase 0.75 (the RV difference between components is larger), that is produced because the absorption component  moves far from the main profile, so we measure the continuum like a fiducial emission and (2) the fact that at these phases the emission feature reaches just the level of the continuum. 

We searched for donor spectral features at this spectral region selecting 
a model spectrum from the  library MARCS (http://marcs.astro.uu.se/)  with  $T_{eff}$= 8000 $K$, $\log g$= 3.5 and $Z= Z_{0}$. 
We found 4 spectral features in the CH2 region, that are listed in Table 6. It is clear that the character labeled Lf2  is the best candidate to produce the absorption component. The origin in the donor is consistent with their half-amplitude, $K_{ab}$= 174 $\pm$ 5 km/s given in Table 5. 



Looking at the donor model spectrum mentioned above, we found that Mg\,II 1091 should have $EW=$ 0.0275 nm for a single star. The observed diluted value  at $\Phi_{o}$= 0.70, viz.\, 0.0150 nm, implies
that the donor contributes about 59\% to the total flux at this time.

Having detected the origin of the moving feature in the donor star, we carefully measured the RV of the  He\,I 1083  line, obtaining  $K_{1}$= 56 $\pm$ 7 km/s (Fig.\,9 and Table 5). 
From the phasing, we conclude that the formation of He\,I 1083 occurs in a region around the primary star. The low amplitude suggests a low mass ratio system.

We built a model spectrum for the donor star at each orbital phase in the region of He\,I 1083 assuming: (1) the $FWHM$ of  the feature Lf2 is constant and its $EW$ matches the observed value around phase 0.70. We used this $FWHM$ for all donor lines at all phases, (2) the relative intensity of the four lines listed in Table 6 remains fixed during the cycles, (3) all lines follow the same (relative) variability that the line Mg\,II given by Eq.\,(3). This meant to apply the right veiling factor to each template at each phase, and (4) all lines follow the motion of the donor star given by the fit to the Mg\,II RVs. The result was a set of templates for the donor contribution at each observed orbital phase for segment CH2.

These templates allowed to explain part of the shape variability observed in He\,I 1083  but not all. We find that:  (1) The absorption/emission feature described in Fig.\,7 probably is the result of  blending with the donor spectral line labeled Lf2 in Table 6 (Fig.\,10), (2) our simple model of blending describes relatively well part of the observed line profile variability (Fig.\,11), 
(3)  the spectra at phases 0.54 and 0.57 show a remarkable blue shading, that cannot be attributed to blending with the donor lines,  and might indicate an outflow through the $L_{3}$ point (Fig.\,12).  The strong red shading  visible at  $\Phi_{o}$=  0.94  could  be absorption by a high velocity gas stream falling from the donor to the gainer (Fig.\,12). 
Both interpretations are consistent  with the interacting nature of the binary. 
Afterwards, we divided the observed spectra by the model spectra to get ''unblended" He\,I profiles. This was possible in several cases but in others some structure remained,  especially around phase 0.6 where it was impossible to ''clean" the feature associated to the Lf2 line. In general, the division procedure introduced additional noise to the spectra and was not adequate to improve RVs for He\,I lines or pursuing additional analysis, except for measuring the $FWHM$, 
that was measured manually with the cursor using the ''splot" IRAF package. This parameter shows
two maxima around phases 0.5 and 0.9 and three minima, around phases 0.10, 0.45 and 0.60 (Fig.\,13). We note that the maxima are attained at similar phases that the $FWHM$ of the He\,I 5875 line (Mennickent et al. 2009b). While our observations are limited to $\Phi_{l}$= 0.5-0.7, the optical observations reported by these authors span all the long/orbital cycles.
The large $FWHM$ changes evidence their origin in a complex environment, 
not a simple stellar photosphere.
An asymmetric extended pseudo-photosphere formed around the primary  with material 
injected from the donor is the probable origin. 
 The maximum around secondary eclipse was already explained 
as outflow through the Lagrangian $L_{3}$ point. The alternative interpretation of the gas stream facing the observer at these phases in rejected since: (1) the asymmetry is visible before and after $\Phi_{o}$=  0.5
and (2) the gas stream has no continuum source behind.   On the other hand the $FWHM$ maximum around $\Phi_{o}$= 0.9 is probably the hallmark of the gas stream, that with large positive velocities produces an enhancement of the profile width.  Finally, we note that He\,I 1083 is still visible around main eclipse ($\Phi_{o}$= 0.94 and 0.07), and their
width is minimum at $\Phi_{o}$= 0.10. This is consistent with a partial eclipse (at $\Phi_{o}$= 0.10) of the forming region
that should dominate the 3th and 4th quadrants and then partially escape through the $L_{3}$ point. In other words, the 1st and 2nd quadrants should have less accreted  material than the 3th and 4th ones.
Eventually, the material at the 1st quadrant might have lower rotational velocities (explaining the narrower profiles) and it is even possible that we observe part of a fast rotating primary at the upper wings of the profile of $\Phi_{o}$= 0.07.

\begin{table}
\centering
 \caption{Results of the fits to the radial velocities with simple sine functions of the form $\gamma$ + $K\sin(2\pi(x+\delta)$). $\gamma$, $K$ and $rms$  are in km/s.}
 \begin{tabular}{@{}ccccc@{}}
 \hline   
Line  &$\gamma$ & $K$ & $\delta$& $rms$ \\
\hline
Mg\,II 1092 &-2 $\pm$ 2 & 181 $\pm$ 3&0.00$\pm$0.00 & 10\\
He\,I 1083 C2 &-107 $\pm$ 4 &174 $\pm$ 5 &0.00$\pm$ 0.01 & 14\\
He\,I 1083 &-13 $\pm$ 5 &-56 $\pm$ 7&0.95$\pm$0.02&24 \\
He\,I 1092 & 16 $\pm$ 6 &-42 $\pm$ 7 &0.00 (fixed)   & 14\\
Si\,IV 139   & -44 $\pm$ 19&-64 $\pm$ 26&0.76$\pm$0.07&45\\
Si\,IV 140   &-66 $\pm$ 6&-35 $\pm$ 8&0.81$\pm$0.04&14\\
Al\,III 185    &-63 $\pm$ 3 & -44 $\pm$ 4 &0.85$\pm$0.02&7\\
Al\,III 186    &-49 $\pm$ 4 &-50 $\pm$ 6 &0.86$\pm$0.02&10 \\
L$\alpha$  &17 $\pm$ 7      &-20 $\pm$ 9    &0.79$\pm$0.08&17\\
\hline
\end{tabular}
\end{table}

\begin{table}
\centering
 \caption{Spectral lines detected in a model spectrum for the donor star in the CH2 region. The $EW$ ratio is with respect to the Lf2 component.}
 \begin{tabular}{@{}ccccc@{}}
 \hline   
label &$\lambda$ (nm)  &$EW$ (nm)& $ratio$ &$FWHM$ (nm)\\
\hline
Lf1&1081.39&0.0091&0.327&0.014\\
Lf2&1082.41&0.0173&0.622&0.089\\
Lf3&1083.01&0.0278&1.000&0.099\\
Lf4&1084.68&0.0151&0.543&0.099\\
\hline
\end{tabular}
\end{table}


\begin{figure}
\scalebox{1}[1]{\includegraphics[angle=0,width=8cm]{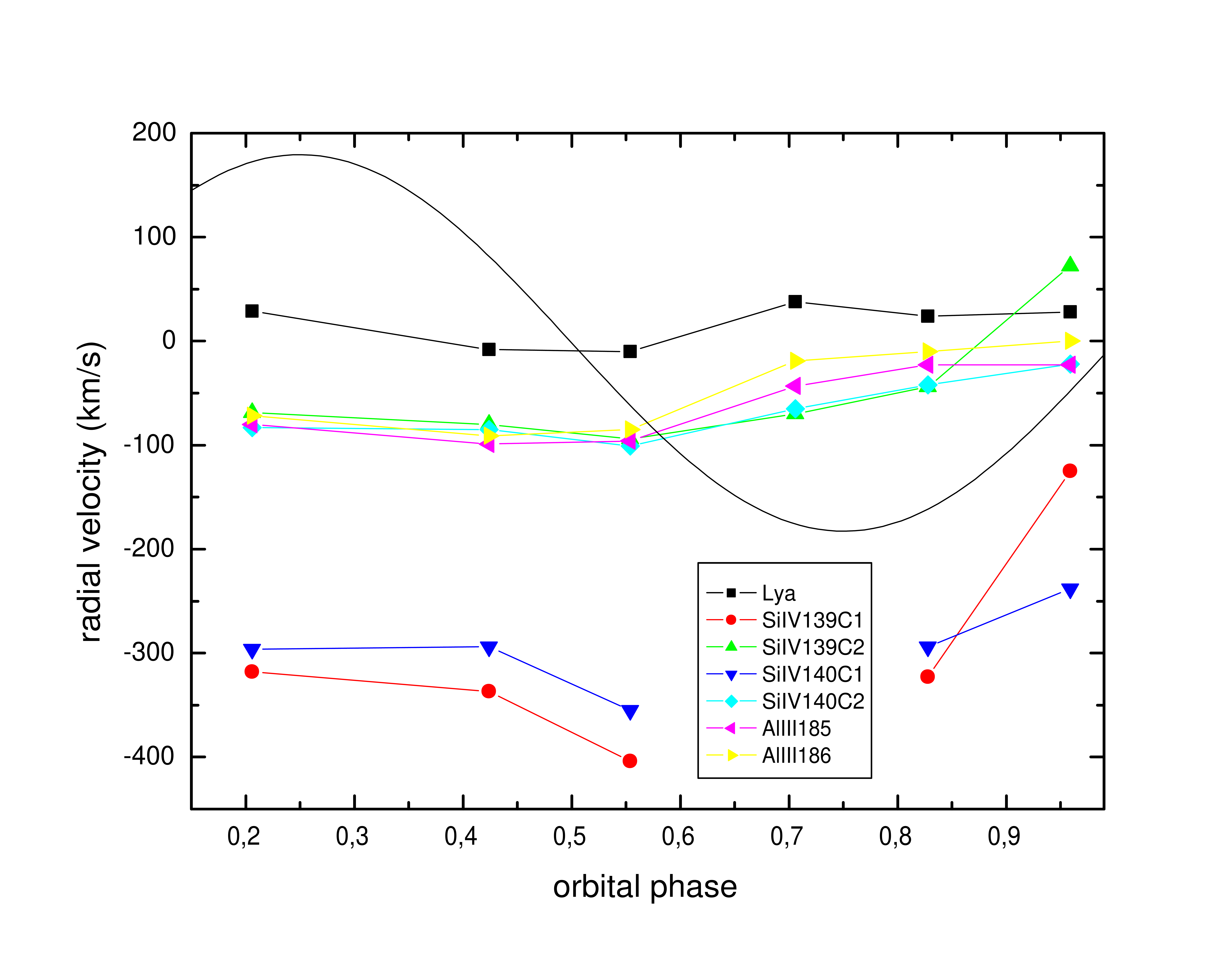}}
\caption{Radial velocity for several lines. The solid line shows the best fit to the donor RV discussed in Section 3.3. C1 and C2 refer to the blue and main components of the asymmetrical lines. A typical error is 7 km/s.}
  \label{x}
\end{figure}

\begin{center}
\begin{figure*}
\scalebox{1}[1]{\includegraphics[angle=0,width=18cm]{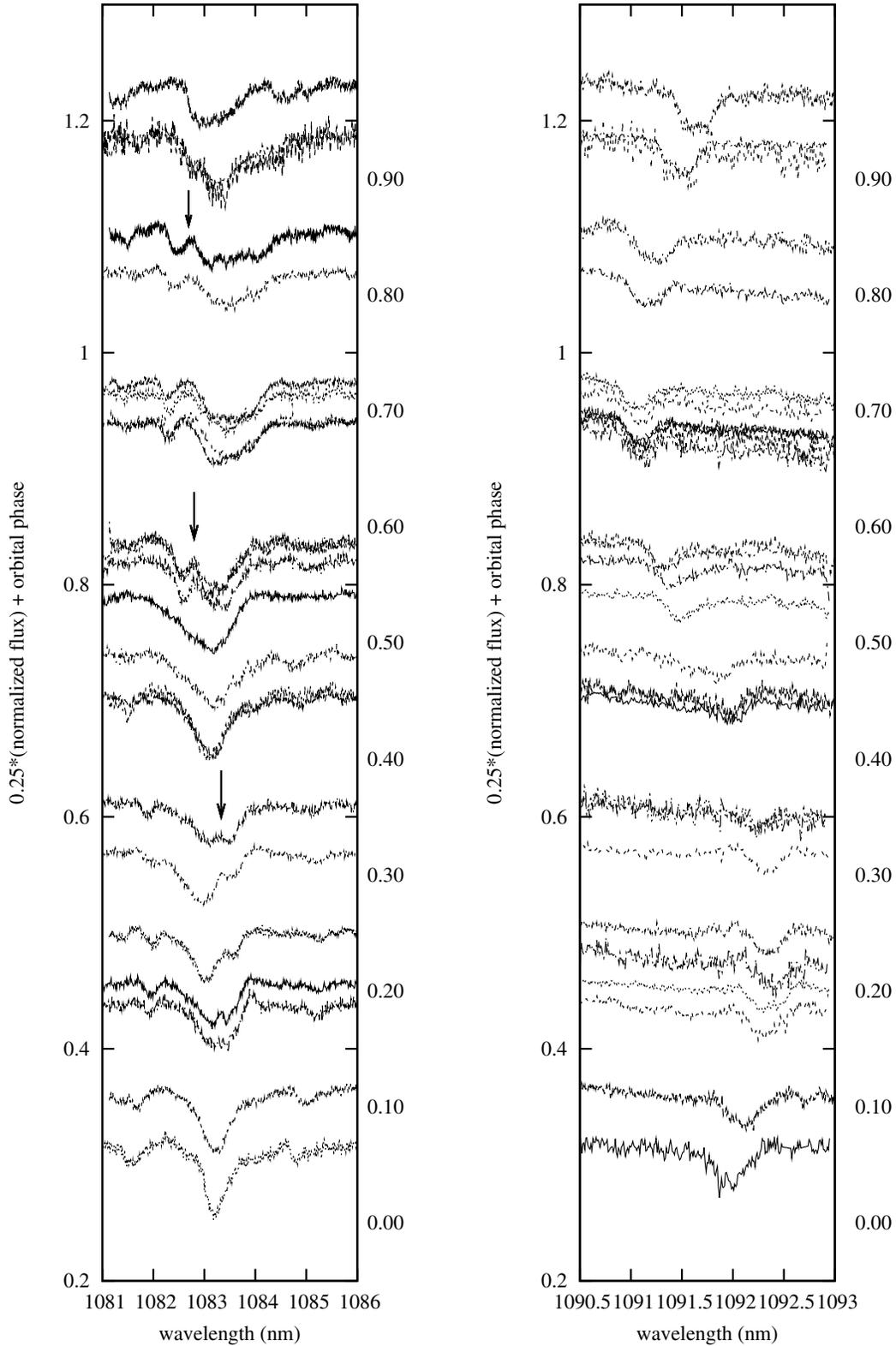}}
\caption{Orbital variability of He\,I 1083 and Mg\,II 1092.
The emission/absorption feature indicated by an arrow
in some profiles is due to contamination by a donor absorption line. 
Labels at the right side of each panel indicate $\Phi_{o}$.}
  \label{x}
\end{figure*}
\end{center}

\begin{figure}
\scalebox{1}[1]{\includegraphics[angle=0,width=8cm]{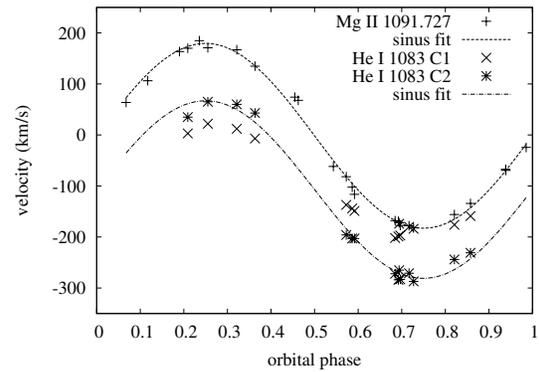}}
\caption{Radial velocity for  Mg\,II 1092 and the best sinus fit, along with velocities for the
He\,I 1083 emission component (C1, probably an artifact) and absorption component (C2, probably the real feature, an absorption from the donor star) indicated in Fig.\,7. Here the velocities of C1 and C2 are tentatively referred to $\lambda$ 1083.306 nm. }
  \label{x}
\end{figure}

\begin{figure}
\scalebox{1}[1]{\includegraphics[angle=0,width=8cm]{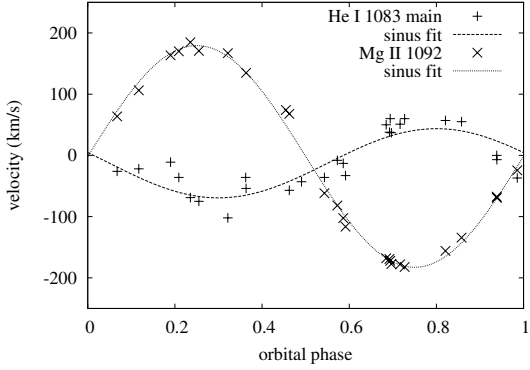}}
\caption{Radial velocity  for Mg\,II 1092 and the best sinus fit, along with velocities for the
He\,I 1083 main absorption component and sinus fit. }
  \label{x}
\end{figure}

\begin{figure}
\scalebox{1}[1]{\includegraphics[angle=0,width=8cm]{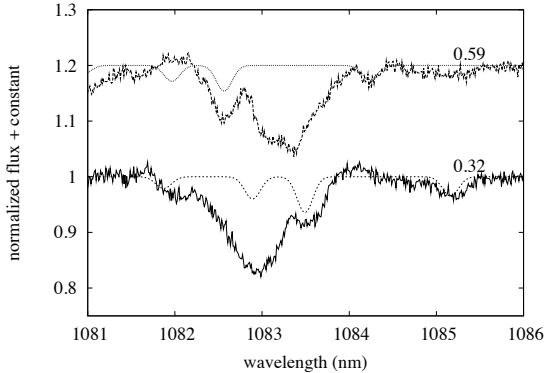}}
\caption{The absorption/emission feature described in Fig.\,7 is the result of  blending with the donor spectral line labeled Lf2 in Table 6.}
  \label{x}
\end{figure}

\begin{figure}
\scalebox{1}[1]{\includegraphics[angle=0,width=8cm]{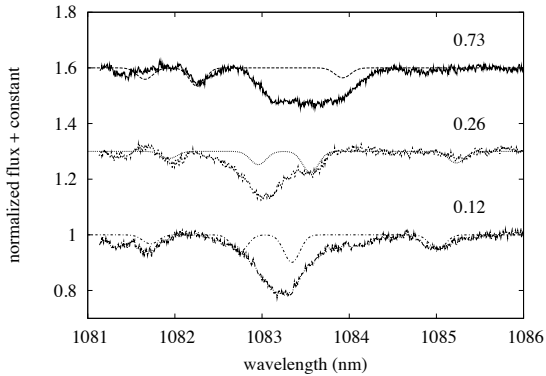}}
\caption{Our simple model of blending describes relatively well part of the observed line profiles.}
  \label{x}
\end{figure}

\begin{figure}
\scalebox{1}[1]{\includegraphics[angle=0,width=8cm]{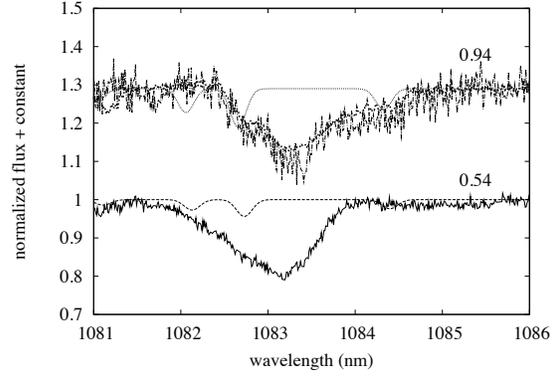}}
\caption{The asymmetries observed in He\,I  spectra at $\Phi_{o}$= 0.54 and 0.94 cannot be explained only by blending with donor features.  Mass flows is a possible interpretation.}
  \label{x}
\end{figure}

\begin{figure}
\scalebox{1}[1]{\includegraphics[angle=0,width=8cm]{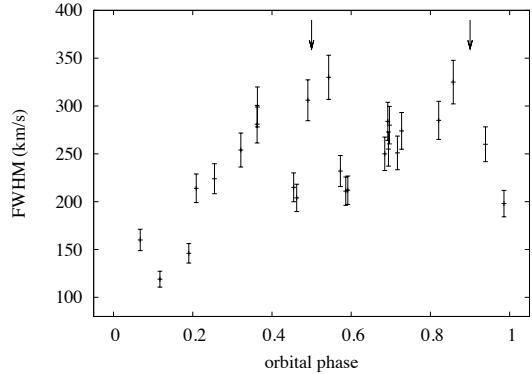}}
\caption{The $FWHM$ of the He\,I 1083 nm  line. Arrows indicate phases where maximum $FWHM$ of He\,I 5875 \AA~ was reported (Mennickent et al. 2009b).}
  \label{x}
\end{figure}

\begin{figure}
\scalebox{1}[1]{\includegraphics[angle=0,width=8cm]{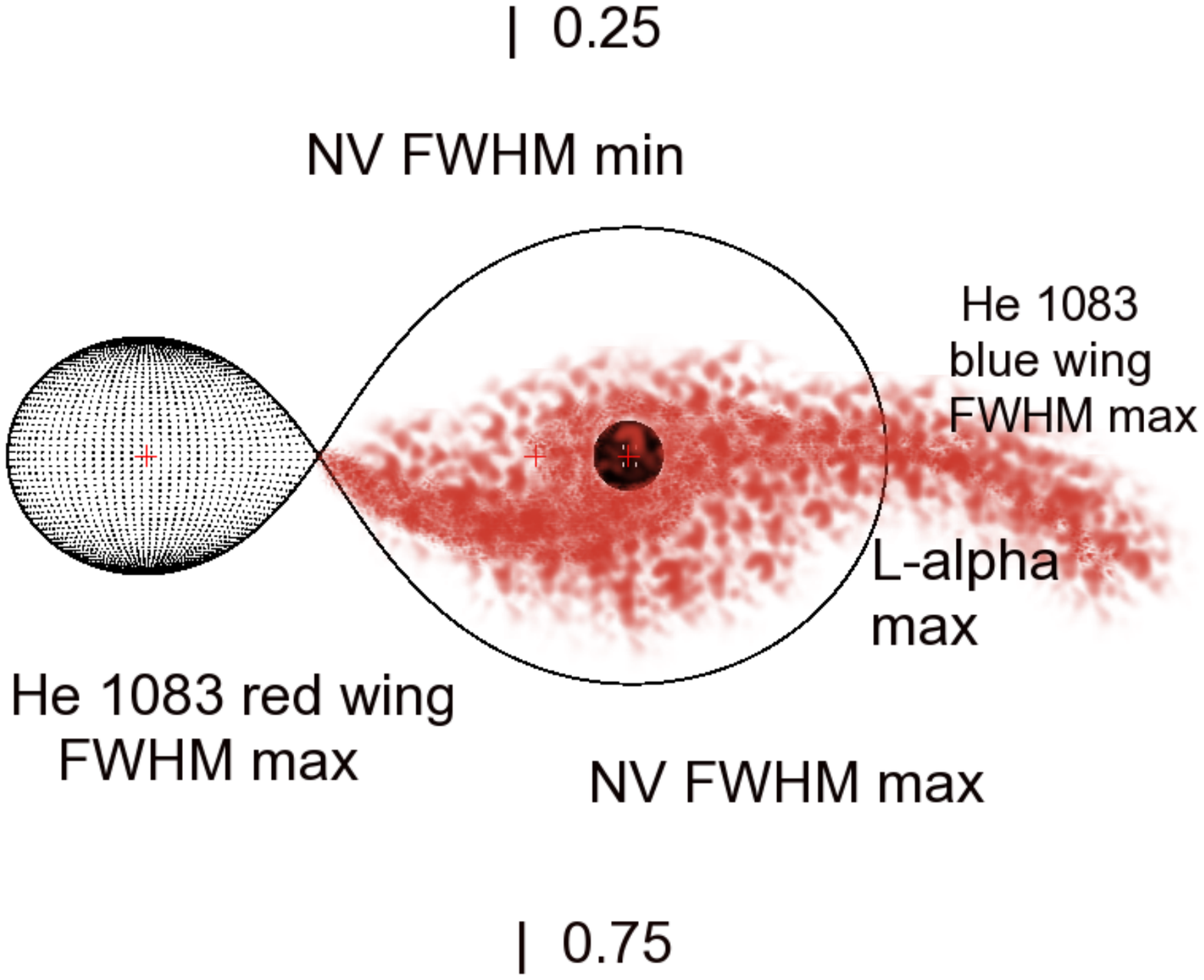}}
\caption{ A sketch of V\,393 Sco and their equatorial mass flow showing some remarkable features observed at specific phases during long cycle minimum ($\Phi_{l}$= 0.49-0.70).  Stellar sizes are scaled according to parameters given in Table 7. }
  \label{x}
\end{figure}

\section{Discussion}

\subsection{The nature of V\,393 Sco and system parameters}

There are several observational features indicating that V\,393 Sco is an interacting binary: the light curve with almost connected minima indicating proximity of the components, the presence of emission in L$\alpha$, Balmer and He\,I lines (M09) and the complex structure of He\,I  1083.

The system mass function for a binary in a circular orbit can be expressed as:\\

$f  = \frac{M_{2}sin^{3}i}{q(1+q)^{2}} = 1.0361\times10^{-7} (\frac{K_{2}}{km s^{-1}})^{3} \frac{P_{o}}{day}$ M$_{\sun}$, \hfill(4) \\

\noindent where $K_{2}$ is the half--amplitude of the RV of the donor star
and $P_{o}$ the orbital period.  The $f$ value derived from Eq.\,(4) and from our radial velocity study 
 is  4.76 $\pm$ 0.24 M$\odot$. 
The $K_{1}$ weighted average including
values of  He\,I 1083, He\,I 1092, Si\,IV 140, Al\,III 185 and Al\,III 186 lines is 45.6 km/s with a probable error of 10\%.
If this  figure reflects the gainer binary motion, then the mass ratio should be $q= M_{2}/M_{1}$= 0.24 $\pm$ 0.02. Both $K$ values yield $a_{2}\sin\,i$=  27.8 $\pm$ 0.5 $R_{\sun}$, $a_{1}\sin\,i$=  6.7 $\pm$ 0.7 $R_{\sun}$, $M_{1} \sin^{3} i \sim$ 7.4 $M_{\sun}$ and  $M_{2} \sin^{3} i \sim$ 1.8 $M_{\sun}$.
 For any possible inclination for this eclipsing system, the mass and temperature for the gainer are consistent with the B3 spectral type estimated in Section 3.2.

We notice that the RV amplitudes of some optical He\,I lines give a possible different $q$ value (viz.\,0.41, M09). However, while $q$= 0.24 yields a stellar mass consistent with the donor temperature of 7250 $K$, the larger $q$ yields $M_{2}\approx 3.0 M_{\sun}$ and a temperature (for a main sequence star) of about 10000 $K$, in conflict with the spectral type derived from optical lines.

Assuming that the secondary star fills their Roche lobe, their mean density should be constrained by the orbital period for
$q \leq$ 0.8 (e.g. Frank, King and Raine 2002):\\

$\overline{\rho} \approx 110 (\frac{P_{o}}{hr})^{-2}$ g cm$^{-3}$. \hfill(5)\\

\noindent In the case of \var we find $\rho$ = 3.2 $\times$ 10$^{-3}$ g cm$^{-3}$ or $\log \rho/\rho_{_{\sun}}$ = $-$2.49. 
 Approximating $R_{2}$ with  the Roche lobe radius of the secondary star (Eggleton 1983):\\

$\frac{R_{2}}{a} =\frac{0.49q^{2/3}}{0.6q^{2/3}+ln(1+q^{1/3})}$   , \hfill(6) \\

\noindent we found $R_{2}$ = 0.265$a$  for $q$ = 0.24. 

 The radius of  the donor  $R_{2}$ $\approx$ 9 $R_{\sun}$ (assuming $i$= 79 degree) indicates an oversized an evolved star. On the other hand, the radius for the gainer $R_{1}$ $\approx$ 2.5 $R_{\sun}$ is smaller than expected 
for a B3 main sequence star. This inconsistency could reflect the limitation of our method when dealing with a complex multicomponent system, as argued in Section 3.1.

Our system parameters corresponds to a much massive and hotter binary  than those given by BD80 and SK90. While these authors gives 
4.2 $M_{\sun}$ and 1.68 $M_{\sun}$, 
and  3.0 $M_{\sun}$ and 1.35 $M_{\sun}$, respectively, our estimates are 7.8 $M_{\sun}$ and  1.9 $M_{\sun}$ for the gainer and donor stellar masses. While SK90 give spectral types of B9 and [F2], ours are significantly hotter, B3 and F0. Unfortunately, we have no access to the source papers to evaluate the work of  the aforementioned authors and to look  for a possible cause of discrepance. However, according to calculations of gas stream dynamics in semidetached mass-exchange binaries (Lubow \& Shu 1975, see also Plavec 1989), the mass ratio and primary radius given by BD80 and SK90, about 0.4 and 0.14$a$ respectively,  do not allow the formation of an accretion disk around the primary, since the gas stream should impact the star. Our optical spectra of V\,393\,Sco (M09) suggests the existence of such a disc (at least at some epochs) and therefore argue against this ``large-$q$" solution.

\subsection{Mass transfer, outflows and mass loss}

Primary evidence for gas flows and interaction region comes  from the observation of L$\alpha$ emission and UV absorption lines. All these lines follow sinusoidal motions consistent with an origin inside the Roche lobe of the primary star, but their half-amplitudes and asymmetries probably indicate different structures. The very low amplitude of the L$\alpha$ emission indicates an origin close to the center of mass of the system, while the larger amplitude of the superionized absorption lines indicate an origin in the primary itself or the pseudo-photosphere responsible of the He\,I absorption (note the similar half-amplitude and phasing for these lines). The L$\alpha$ emission orbital curve is very asymmetrical, rising rapidly to maximum just after secondary eclipse and then declining smoothly.
This behavior is consistent with an origin near the site where the stream from the secondary star hits the circumprimary envelope.   We note that the low mass ratio allows, in principle,  the formation of a disc around the primary.  The fact that 
maximum approaching velocity is reached just after $\Phi_{o}$= 0.5 for this feature, is consistent with the view that at this phase we observe the gas stream almost completely projected on the sight line,
with large negative velocity.
This stream is usually deflected in the direction of the orbital motion due to Coriolis forces, dominating the third and fourth  quadrant, visible just after secondary eclipse,  but it could also splash into larger latitudes producing a region where the observed L$\alpha$ emission arises.  Consistent with this view is the small L$\alpha$  $FWHM$, indicating a region not dominated by the rapid rotation found in the orbital plane. Another
possible origin, material escaping through the L3 point at large angles and not projected against the gainer, is rejected due to the small RV amplitude of the feature. 

We note that the blue components of the superionized lines show larger half-amplitudes than the L$\alpha$ emission and a notable $\gamma$ shift, indicating probably a hot wind emanating from the primary star with projected velocities of 300 or 500 km/s. 
On the other hand, in the IR we observe asymmetries  in the He\,I 1083 line before secondary eclipse indicating equatorial  outflow velocities of the order of 500 km/s. Evidence for this outflow  was discussed in detail in section 3.4. While the outflow derived from this line is strongly phase dependent, indicating motion of material in the orbital plane, probably gas escaping through the Lagrangian $L_{3}$ point, the UV outflows have a comparatively weaker dependence on the orbital phase, and could indicate motion of material above the orbital plane at intermediate or large latitudes. Our conclusion that the outflow seen in He\,I lines at $\Phi_{o}$= 0.5 occurs in the equatorial plane seems to be reinforced by the fact that Si\,IV lines do not show asymmetry at this phase.
The different ionization temperature  for He\,I and UV superionized lines  also suggests a different origin  for these outflows. As the binary is eclipsing, the true velocities for the hotter outflow should be
much larger than the projected velocity of 500 km/s. As these velocities are much larger than the escape velocity of the system ($\sim$ 30 km/s), we conclude than the binary is loosing matter into the interstellar medium by polar and equatorial outflows. 

 The strong asymmetry observed around $\Phi_{o}$= 0.5, comparable in strength to that observed near $\Phi_{o}$= 0.9, indicates a considerable equatorial mass loss rate in the system, even comparable to the mass transfer rate. This result suggests that spectroscopy at $\Phi_{o}$= 0.5 sampling the long cycle might yield valuable diagnostic tools for understanding cyclic mass loss in this object. A sketch of V\,393 Sco and their equatorial mass flow is given in Fig.\,14.


\begin{table}
\centering
 \caption{Summary of parameters for V\,393 Sco. All data is from this paper except the orbital ephemeris from Kreiner (2004).  }
 \begin{tabular}{@{}cc@{}}
 \hline   
Parameter  &value \\
\hline
Ephemeris$_{min, o}$                & 2\,452\,507.7800 + 7.7125772 $\times\,E$  \\
Ephemeris$_{max, l}$                &2\,452\,520.00 + 255.0$\times\,E$    \\
$E(B-V)$                  &0.13 $\pm$ 0.02    \\
$d$                           &520 $\pm$ 60 pc  \\
$f(M)$                      &4.76 $\pm$ 0.24 $M_{\sun}$  \\
$a_{2}\sin\,i$            &27.8 $\pm$ 0.5 $R_{\sun}$ \\
$a_{1}\sin\,i$             &6.7 $\pm$ 0.7 $R_{\sun}$  \\
$a\sin\,i$ &34.5 $\pm$ 1.2   $R_{\sun}$  \\
$M_{1}\sin^{3}\,i$    &$\sim$ 7.4 $M_{\sun}$  \\
$M_{2}\sin^{3}\,i$    &$\sim$ $1.8 M_{\sun}$  \\
$q$  & 0.24 $\pm$ 0.02 \\
$T_{1}$    &19000 $\pm$ 500 K \\
$T_{2}$    &7250 $\pm$ 300 K \\
$\log g_{1}$ &4.5 $\pm$ 0.3 \\
$\log g_{2}$ &3.0 $\pm$ 0.3 \\
$R_{1}/R_{2}$ &0.27 $\pm$ 0.03 \\
$R_{2}$ &0.265$a$ \\
\hline    
\end{tabular}
\end{table}

\subsection{Comments on the long cycle}

V\,393 Sco shows a long photometric cycle of 255 days (Pilecki \& Szczygiel 2007). Our infrared spectroscopy span a small fraction of this cycle, only between phases 0.49 and 0.70, so for this dataset is difficult to trace conclusions  on the long cycle. Distinct is the situation for the IUE spectra, although few, they are very well sampled not only in orbital cycle but also in the long cycle as well. We have 7 spectra around $\Phi_{l}$=  0.9 (the maximum), two around $\Phi_{l}$=  0.46 (the minimum) and one at $\Phi_{l}$=  0.28.  The important result is that there is no major change between fluxes and line shape at these epochs; the UV variability is mainly orbital, and large changes
are not  observed with  the long cycle. This is clear when comparing the large fluxes detected at   $\Phi_{l}$= 0.92 (SWP36222) and  $\Phi_{l}$= 0.28 (SWP38632) and the very different fluxes detected at 
 $\Phi_{l}$= 0.93 (SWP36232) and  $\Phi_{l}$= 0.92 (SWP36222). The same is true for the H$\alpha$ emission, that is modulated with the orbital period, not the long cycle.
 
 The above fact must be contrasted with the know issue that long photometric variability is larger in redder bandpasses (M03, Michalska et al. 2009). This suggests that high latitude hot outflows are not directly connected with  the long cycle, and that the long cycle is more related to cooler mass outflows driven across the equatorial plane,  not through the hotter polar regions. An additional point in favor of this view, is the fact that DPVs seen at intermediate latitudes, i.e. those showing ellipsoidal rather than eclipsing variability, do not show larger amplitude 
 long cycles (M03).
 
If matter is being expelled in the equatorial plane into the interstellar medium, a reservoir of gas could be formed around the binary, something like a circumbinary disk. Actually, some evidence for this structure was found in the DPV OLGE\,05155332-6925581 by M08. It is notable that for V\,393 Sco we find no trace of additional infrared sources in our SED model. This could be a real feature of the system, but could also be due to  oversimplification in our model, since we have not considered the circumprimary matter.

\subsection{Comparison with  the UV spectra of Algols \& Serpentids and final remarks}

A narrow L$\alpha$ emission core, as seen in V\,393 Sco, is also visible in the Algol TT\,Hya (Miller at al. 2007). However, we note the absence of Fe\,II lines in the IUE spectra of \var, especially in the Al\,III line region; this contrasts with the observation of TT Hya  (and $\beta$ Per, Wecht 2006) and could be due to a hotter pseudo-photosphere in V\,393 Sco  compared with those of classical Algols. In addition, superionized emission lines are not observed in V\,393 Sco, in contrast with the W Serpentid type stars RY\,Per (in this star they are observed near totality, Olson \& Plavec 1997) and $\beta$ Lyr (Aydin et al. 1988).  Finally, the UV lines detected in V\,393 Sco are similar to those observed in other Galactic DPV, AU Mon (Sahade \& Ferrer 1981). All these observations point to hot regions with different temperatures in 
Algols, DPVs and W Serpentids, and support
the view that these classes are interacting binaries with different mass transfer rates regimes (Mennickent et al. 2009b). An exhaustive comparative analysis of all available UV spectra for Algols,
DPVs and  W Serpentids is needed to confirm this point, but it is beyond the scope of this paper. 

One important question to rise here is why V\,393 Sco looses matter into the interstellar medium whereas others low $q$ Algols apparently do not. In all low $q$ semi-detached Algols accretion can form a disk around the gainer. M08 speculated  that in DPVs the mass transfer rate $\dot{M}$ is so large than the primary rapidly spin up until critical rotation, and material cannot be longer be accreted onto the primary, starting to escape from the system through places of minimum binding energy, like the $L_{3}$ point. How the long cycle is produced, and the question if it represents cycles of enhanced mass loss as proposed by M08, still are matters of research, but precession of the circumprimary envelope or
some rotational instability at the primary have been invoked as possible causes (Mennickent et al. 2009a, Mennickent et al. 2009b).  Alternatively, for the DPV AU\,Mon, Peters (1994)  interpreted the long cyclic activity as due to variable mass
transfer caused by a slow pulsation of the secondary about its Roche
surface. An open question is how the hitherto transient radio flares detected in V\,393 Sco are related with the long cycle. Radio flares are probable indicators of activity in the secondary star (e.g. Retter, Richards \& Wu 2005). Evidently, more research at these wavelengths is needed to have a complete picture of the role of the secondary star in the long cycles of V\,393 Sco.

\section{Conclusions}

Based on a study of  infrared and UV spectra of V393 Sco, along with published multiwavelength broad band photometry, we conclude:\\ 

\begin{itemize}
\item  The fit to the SED  of V\,393 Sco with two stellar models allowed to calculate the stellar temperatures, surface gravities, color excess and distance that are given in Table 7.
Our study of RVs of infrared He\,I, Mg\,II  lines and UV resonance lines yielded a mass ratio $q$= 0.25. 
\item Based on the larger distance ($d$= 523 $\pm$ 60 pc) and reddening ($E(B-V)$= 0.13 $\pm$ 0.02), we argue that V\,393 Sco  is not a member of the cluster M7. 
\item The blue depression observed in He\,I 1083 profile  around secondary eclipse suggests the existence of mass loss through the Lagrangian $L_{3}$ point with velocities $\sim$ 300 km/s. This material probably does not return to the system and it is lost into the interstellar medium. 
\item  The He\,I 1083 asymmetry around $\Phi_{o}$= 0.5 is comparable in strength to the asymmetry observed around $\Phi_{o}$= 0.9 (a diagnostic of the gas stream), suggesting that a large fraction of the accreted matter is lost into the interstellar medium during $\Phi_{l}$= 0.49-0.70.
\item The general variability of He\,I 1083, especially the width changes, indicates the presence  of a hot optically thick envelope or pseudo-photosphere that mimics the appearance of a hot B-type primary. We argue that this region is asymmetrical and dominates the 3th and 4th quadrants. The irregularities observed in the He\,I profile during the orbital cycle suggests that matter in this circumprimary envelope posses a complex dynamics, and is not rotating at Keplerian velocities around the primary.
\item As circumprimary matter was not included in our SED model, caution must be taken when considering the derived stellar parameters. 
\item Asymmetries seen in superionized UV lines at almost all observed orbital phases point to the existence of a permanent hot wind at high latitudes. This wind is associated to the hotter star, has velocity $>$ 500 km/s and is an important mass loss channel for the binary.
\item Contrary to W\,Ser stars, we report the absence of high ionization emission lines in the UV spectrum of V\,393 Sco. The UV spectra of W\,Serpentids, DPVs and Algols could reveal distinctive features.
\item The low amplitude of UV continuum/line variations during long cycle,  argue against polar jets as the cause for the long-term variability. The long-term variability is probably related to equatorial mass loss.
\item Two additional arguments supporting this view are: (1)  the similitude between long-term amplitudes observed in non-eclipsing and eclipsing DPVs by M03 and (2) the reported larger amplitude in red bandpasses for  the long cycles.  
\end{itemize}

\section{Acknowledgments}

 We acknowledge the anonymous referee for useful comments and suggestions on the first version of this paper. We thank Dr. Geraldine Peters for useful discussions concerning IUE data during the preparation of this paper. More importantly, we thank her for carry out the observational project that provided the available IUE data for V\,393 Sco that is discussed here.  
REM acknowledges support by Fondecyt grant 1070705, the Chilean 
Center for Astrophysics FONDAP 15010003 and  from the BASAL
Centro de Astrof\'isica y Tecnologias Afines (CATA) PFB--06/2007.
This publication makes use of data products from the Two Micron All Sky Survey, which is a joint project of the University of Massachusetts and the Infrared Processing and Analysis Center/California Institute of Technology, funded by the National Aeronautics and Space Administration and the National Science Foundation.

\bsp 
\label{lastpage}
\end{document}